\documentclass[fleqn,usenatbib,useAMS]{mnras}

\usepackage{newtxtext,newtxmath}

\usepackage[T1]{fontenc}

\DeclareRobustCommand{\VAN}[3]{#2}
\let\VANthebibliography\thebibliography
\def\thebibliography{\DeclareRobustCommand{\VAN}[3]{##3}\VANthebibliography}

\usepackage{graphicx}
\usepackage{amsmath}

\newcommand{\logMdisc}{$\log(M_{\text{disc}}/M_{\sun})$}
\newcommand{\logRcrit}{$\log(r_{c}/\text{AU})\ $}

\title[The Birth and Future of HD 163296]{The CO-Fuelled Time Machine: Tracing Birth Conditions and Terrestrial Planet Formation Outcomes in HD 163296 through Pebble Drift-induced CO Enhancements}

\author[Williams and Krijt]{
Joe Williams,$^{1}$\thanks{E-mail: jw1436@exeter.ac.uk}
Sebastiaan Krijt,$^{1}$
\\
$^{1}$School of Physics and Astronomy, University of Exeter, Stocker Road, Exeter EX4 4QL, UK\\
}

\date{Accepted ---. Received ---; in original form ---}

\pubyear{2025}

\begin{document}
\label{firstpage}
\pagerange{\pageref{firstpage}--\pageref{lastpage}}
\maketitle

\begin{abstract}

The architecture and composition of planetary systems are thought to be strongly influenced by the transport and delivery of dust and volatiles via ices on pebbles during the planet formation phase in protoplanetary discs. Understanding these transport mechanisms is crucial in building a comprehensive picture of planet formation, including material and chemical budget; constraining the birth properties of these discs is a key step in this process. We present a novel method of retrieving such properties by studying the transport of icy pebbles in the context of an observed gas-phase CO enhancement within the CO snowline in the protoplanetary disc around HD 163296. We combine Markov Chain Monte Carlo (MCMC) sampling with a fast model of radial drift to determine the birth gas mass and characteristic radius of the disc, and compare our results against observations and models in the literature; we find the birth-condition disc gas mass to be $\log_{10}(M_{\rm{disc}}/M_{\sun})=-0.64^{+0.19}_{-0.24}$ and the characteristic radius to be $\log_{10}(r_{\rm{c}}/\rm{AU})=2.30^{+0.45}_{-0.46}$. We additionally determine that dust grains must be `fragile' ($v_{f}=100~\mathrm{cms}^{-1}$) to retain enough dust to match current dust mass observations, with our lowest fragmentation velocity model providing a current-age dust mass of $\rm{M_{dust}}=662^{+518}_{-278} \rm{M_{\earth}}$ based on the retrieved birth conditions. Using our retrieved birth conditions, we extend our simulations to mass of material reaching the water snowline in the inner disc, where terrestrial and super-Earth planets may be forming, and speculate on the nature of these exoplanets.
\end{abstract}


\section{Introduction}

Understanding the full process of planet formation is a challenging task, considering the vast size and mass scales covered in astronomically short time scales of a few million years. Overcoming this time constraint can be challenging for classical planet formation models, such as core accretion \citep{dodson-robinson2009, lambrechts&johansen2012, johansen&bitsch2019}, although this framework has demonstrated success in building both terrestrial planets \citep[e.g.][]{raymond2006} and giant planets \citep[e.g.][]{emsenhuber2021}. Alternative modern planet formation theories, such as pebble accretion, are capable of rapidly creating the cores of gas giants in less than a million years \citep[e.g.][]{johansen&lambrechts2017_review}; these rapid-formation models are underpinned by the radial drift and accretion of `pebbles', which are key in creating terrestrial bodies and gas giants within the required timeframes \citep{lambrechts&johansen2012, ormel2017, drazkowska_pp7_2023}. Relatively recent models look to combine both pebble and core accretion  \citep[e.g.][]{alibert2018_hybrid_accretion}, demonstrating the importance of each formation theory.

The underlying pebbles in pebble accretion theory typically range in size from a few millimetres to around a meter, although they are best defined through their Stokes number, which describes the strength of their coupling to the gas in the protoplanetary disc \citep{ormel2017, drazkowska_pp7_2023}; particles with lower Stokes numbers are well-coupled to the gas and are, generally speaking, smaller in size. As pebbles grow, they begin to decouple from the gas as their Stokes number increases and drift radially inwards to the central star due to drag from gas \citep{weidenschilling1977, liu&ormel2018}. Most particles are expected to drift due to growth \citep{birnstiel2012, stammler2023}. This radial drift mechanism is thought to be vital in transporting material for the growth of planetary embryos, offering faster growth mechanisms compared to traditional models of planet formation \citep{lambrechts&johansen2012, johansen&lambrechts2017_review, drazkowska2021}. 
The inner system's architecture and properties also depends on the amount of pebble drift through the system \citep{lambrechts2019}. The emerging embryos and resultant planets will in turn interact with both the disc and each other, impacting the system architecture further as well as the pebble flux from the outer disc by way of gaps \citep{rice2006, paardekooper&mellema2006, zhu2012, baruteau2014, pinilla2016, kalyaan2021, stammler2023}. Understanding the complex relationship between forming planets and their disc is challenging, but reveals much about the formation processes and final outcomes.

Pebble drift as a means of material transport is also crucial to the effective delivery of different volatiles from the outer disc to the inner regions \citep[e.g.][]{cuzzi&zahnle2004, krijt2018, bosman2018, booth&ilee2019, kalyaan2021}; studying this transport as well as the snowlines of particular volatiles is vital in understanding the composition of forming planets, as well as providing the material in the first place \citep{oberg2011, oberg2016, oberg&bergin2021}. Studying the molecular abundances and emission line intensity profiles of molecular tracers in discs can hence be an incredibly informative probe of pebble drift, and so forms a useful tool in exploring the properties of the disc and any forming planets, as well as linking the chemical reservoirs of the outer disc to the inner regions. \citet{zhang2020} observed three CO isotopologues ($^{13}$C$^{18}$O ($2-1$), C$^{17}$O ($2-1$), and C$^{18}$O ($2-1$)) in the protoplanetary disc around HD 163296 with the NOEMA interferometer and matched the observed line spectra with thermochemical models using models from \citet{zhang2019}. They found that the CO abundance within the CO snowline ($\lesssim 70$ AU) is likely to be between 1.8 and 8 times higher than expected, which they used to estimate that a time-integrated mass flux of 150-600 $M_{\earth}$ must have drifted within the CO snowline (assuming a disc age of 5-10 Myr). This pebble flux estimate is likely a lower limit, as the the actual C/H enhancement may be higher than observed due to an overestimated disc mass \citep{zhang2020}. If the enhancement is larger, then a higher pebble flux would be required. Studying the transport mechanisms that delivered this mass of CO to the inner disc can help build a picture of the disc and the conditions that caused this pebble drift, and therefore allow us to infer fundamental properties of the disc itself, such as its mass.

Estimating the mass of protoplanetary discs is essential to calculating the available material and chemical budget for planet formation \citep[e.g.,][]{drazkowska_pp7_2023, krijt_pp7_2023}. Unfortunately, determining disc mass is a challenging task, as cold molecular hydrogen cannot generally be observed directly. Instead, gas mass estimates typically rely on using other molecular tracers such as CO \citep[e.g.,][]{bergin&williams2017}. Using different molecular tracers and assumptions in model-based inference when determining disk mass can then lead to very different results, as illustrated for example in Fig.~10 of \citet{miotello2023}. An alternative approach, proposed by \citet{Powell2017}, relies on using continuum observations to measure the radial extend of the pebble disc at different wavelengths, which can provide constraints on the underlying gas surface density (and hence total mass) when combined with standard dust coagulation plus drift models. This approach has been found to yield relatively high disc masses (between 9-27\% of the stellar mass) for a handful of nearby discs including HD 163296 \citep{Powell2019}. Since HD 163296 may contain planets \citep{teague2018, teague2021, pinte2018, guidi2018, izquierdo2022}, these additionally can be used as a means to measure disc masses using planet migration mechanisms in tandem with dust dynamics \citep[][]{wu2023}.

In this paper we explore a new method to determine the initial gas mass of the disc at birth (the start of the class-II stage): one that relies on (a) knowing the cumulative radial flux of solids through a particular snowline, and (b) models of pebble formation and drift. We focus on the mass flux reported by \citet{zhang2020}, which offers a unique opportunity to study history and conditions of pebble drift in HD 163296's disc, and combine radial drift models with MCMC sampling to explore the posterior distribution of the disc gas mass and characteristic radius. The paper is structured as follows: we introduce our disc model and cumulative pebble flux in section~\ref{sec:disc_model}; we detail our approach with an MCMC sampler in section~\ref{sec:MCMC_set_up}; we introduce a test case of a synthetic disc in section~\ref{sec:test_case}; we extend our methodology to HD 163296 in section~\ref{sec:additional_set_up} and cover the results in section~\ref{sec:HD 163296_results}; we then discusses the assumptions, limitations, and implications of our work in section~\ref{sec:discussion}; and we highlight important conclusions and key results in section~\ref{sec:conclusions}.

\section{Methodology}
\label{sec:methods}

\subsection{Pebble Drift for Material Enhancement}
\label{sec:disc_model}

\begin{figure*}
	\includegraphics[width=0.8\paperwidth]{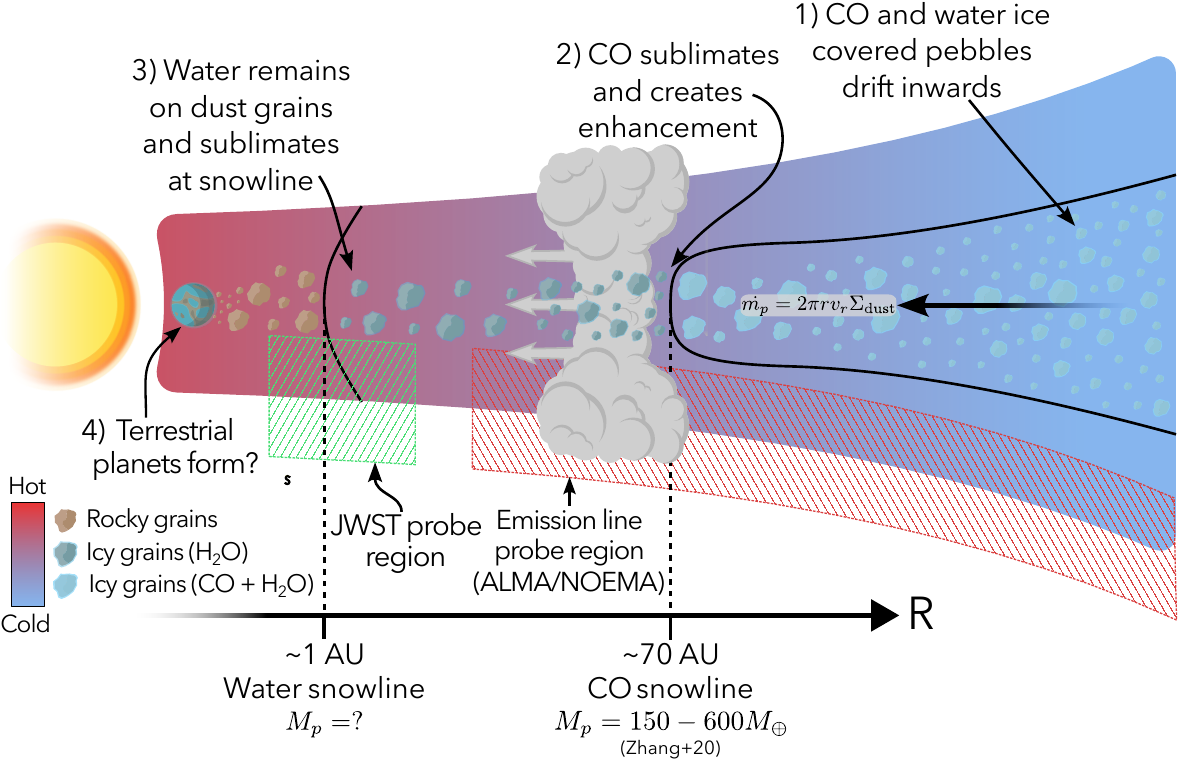}
    \caption{Schematic illustrating transport mechanisms of volatiles that produce the C/H enhancement seen within the CO snowline in HD 163296 as observed by \citet{zhang2020}. The process begins in the outer disc with dust growing to pebbles as the CO and water vapour freeze out onto the pebbles. As the grains grow, they decouple from the gas and drift inwards towards the CO snowline; within the snowline, the CO sublimates and enhances the C/H ratio of the gas and vapour present; the CO vapour is transported to the upper layers of the disc via diffusion. It also advects inwards, indicated by the grey arrows. Water ice remains on the pebbles and is transported further inward as the pebbles continue to drift toward the water snowline, which is around 1 AU in our model in this paper. Here, water sublimates and becomes readily available for planet formation in the region where terrestrial planets may be forming. Note that the illustrated emission line probe region is specific to CO rotational lines in this context. This schematic was inspired by Fig.~2 of \citet{cuzzi&zahnle2004}.}
    \label{fig:transport_schematic}
\end{figure*}

In the context of the work by \citet{zhang2020}, we are investigating the cumulative pebble flux through the CO snowline at 70 AU. To produce the observed enhancement, 150 - 600 $M_{\earth}$ of solid material is required to be delivered through the CO snowline. The processes that deliver this material are diagrammatically shown in Fig.~\ref{fig:transport_schematic}: CO and water in the upper layers of the disc freeze onto grains in the midplane of the outer disc while the grains coagulate, grow, and drift at a rate given by equation~\ref{eqn:pebble flux}. Once the pebbles drift inside the CO snowline, the CO ice sublimates and creates a local enhancement that rises via diffusion into the warm molecular layer, where it is observed. We visualise this vertical transport in the form of a `plume' of vapour (although we note that the CO does not necessarily distribute into the vertical layers in the style of a plume). The grains retain their water and drift further towards the star, then sublimating the ice within the water snowline, where terrestrial planets may be forming.

We are particularly interested then in the time-integrated, or cumulative, pebble flux through a particular location in the disc $r$ at some time $t$. This is readily given by the equation:

\begin{equation}\label{eqn:cumulative flux}
    M_{p}(r,t) = \int_{0}^{t} \dot{m_{p}}(r,t') dt'
\end{equation}
where $\dot{m_{p}}(r,t)$ represents the instantaneous mass flux of pebbles, which can be written as \citep{birnstiel2009, drazkowska2021}:

\begin{equation}\label{eqn:pebble flux}
    \dot{m_{p}}(r,t) = 2\pi r v_{r}(r,t) \Sigma_{\rm{d}}(r,t)
\end{equation}
where $v_{r}$ is the radial drift velocity of the pebbles, and $\Sigma_{\rm{d}}$ the surface density of pebbles. The radial velocity of pebbles depends on particle size (through the Stokes number, see below), which in turn depends on dust coagulation and fragmentation, making the instantaneous mass flux a complex function of time and position in the disc \citep[e.g.][]{birnstiel2012, lambrechts&johansen2014, lambrechts2019, drazkowska2021, bitsch2023}.

\begin{figure}
	\includegraphics[width=\columnwidth]{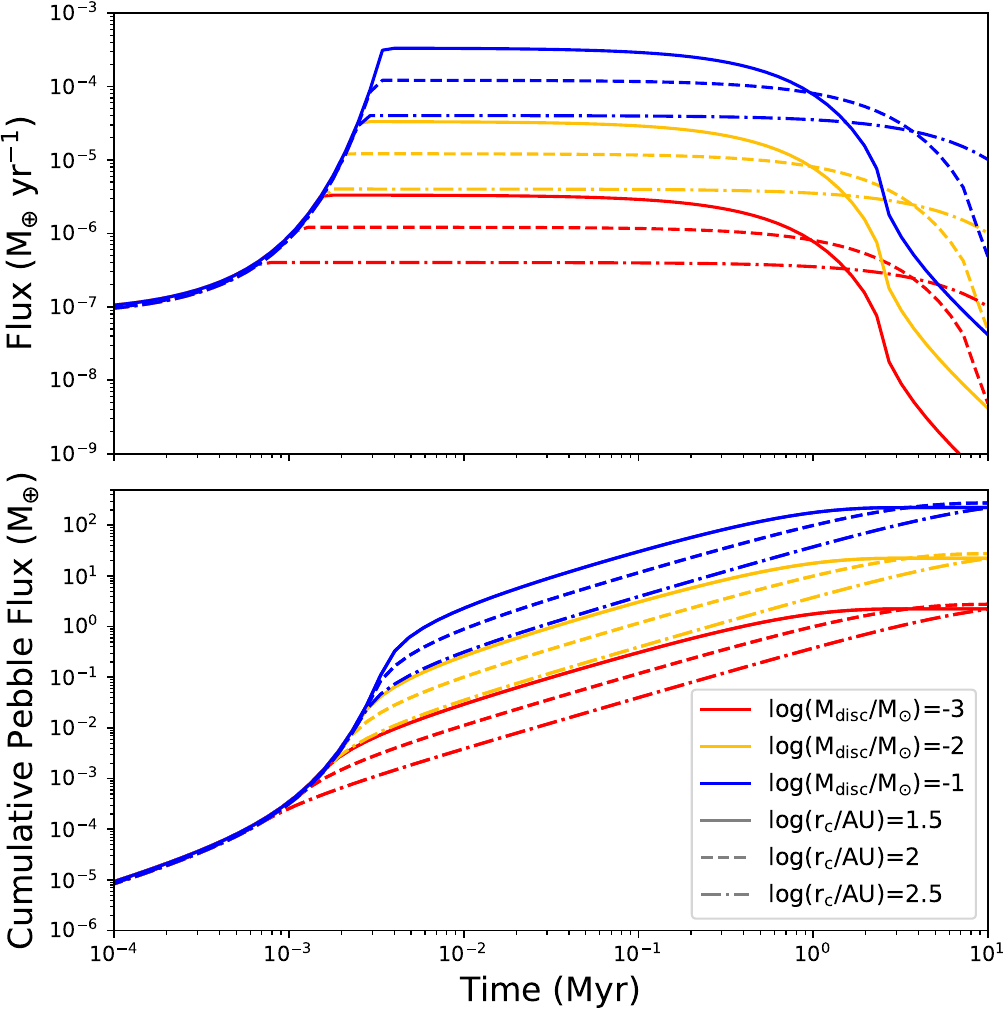}
    \caption{The instantaneous pebble flux at 10 AU (equation~\ref{eqn:pebble flux}) and cumulative pebble flux (equation~\ref{eqn:cumulative flux}) as functions of time for a disc of arbitrary properties, assuming $\alpha = 1\times 10^{-3}$ and $v_{f} = 100 \mathrm{cms}^{-1}$ as calculated by \texttt{pebble predictor}. The colour of the curves correspond to discs with different masses (red for \logMdisc$=-3$; yellow for \logMdisc$=-2$; and blue for \logMdisc$=-1$) and the line style corresponds to different characteristic radii (solid for \logRcrit$=-1.5$; dashed for \logRcrit$=-2$; and dot-dashed for \logRcrit$=-2.5$).}
    \label{fig:PP_examples}
\end{figure}

Examples of the instantaneous and cumulative pebble flux through 10 AU over time for an arbitrarily defined disc can be seen in Fig.~\ref{fig:PP_examples}. These were calculated using the methodology outlined in Sect.~\ref{sec:discpebblemodel}. Each curve illustrates a disc a different gas disc mass and characteristic radius. It is clear that massive discs (blue) produce a higher mass flow rate and overall cumulative pebble flux at 10 AU; discs of the same mass but larger radial extent see a lower maximum instantaneous flux due to a lower dust surface density $\Sigma_{\rm{d}}$ at 10 AU, as well seeing as the instantaneous flux tapering off more slowly as the discs age. Discs of the same mass but different size accumulate approximately the same pebble flux by 10 Myr, with larger discs winning out since they contain more mass exterior to 10 AU; however, each disc exhibits approximately the same behaviour in the first $\sim$1,000 years (i.e. a few pebble growth timescales) before quickly diverging as the different drift limits take effect. It is worth noting that as the measurement position (10 AU here) increases to much larger values (e.g. 200 AU), then less massive yet physically bigger discs tend to exhibit both a larger instantaneous and cumulative flux than massive compact discs. 

\subsection{Disc and Pebble Drift Model}\label{sec:discpebblemodel}

In order to investigate the co-evolution of pebble growth and drift, we use the 1D simulator \texttt{pebble predictor}\footnote{\texttt{https://github.com/astrojoanna/pebble-predictor}} as introduced by \citet{drazkowska2021}. \texttt{pebble predictor} is capable of quickly computing pebble drift mass fluxes, solving for pebble growth, fragmentation, and radial drift in a static gas disc with a static temperature profile. It does not include the formation and effects of substructures within the disc or consider planetesimal formation and accretion by default. We consider these limitations in detail in section~\ref{sec:assumptions_and_limitations}. 

We briefly reiterate the main components of \texttt{pebble predictor} relevant to this work, and refer the reader to \citet{drazkowska2021} for more details. The static gas disc follows the equation:

\begin{equation}\label{eqn:gas profile}
    \Sigma_{\rm{g}}(r)= \frac{(2-\gamma)M_{\rm{disc}}}{2 \pi r_{c}^{2}} \left( \frac{r}{r_{c}} \right)^{-\gamma} \exp \left[- \left(\frac{r}{r_{c}} \right)^{2-\gamma} \right]
\end{equation}
with disc mass $M_{\rm{disc}}$ and characteristic scaling radius $r_{c}$ \citep{zhang2021, andrews2011}. The power-law index $\gamma$ is often set to 1 \citep[e.g.][]{andrews2009, andrews2011}. Similarly to the surface density, the midplane temperature profile remains fixed according to the equation:

\begin{equation}\label{eqn:temperature profile}
    T(r) = T_{0} \left(\frac{r}{\rm{AU}} \right)^{-0.5}
\end{equation}
where $T_0$ is a scaling factor.

For the solids, \texttt{pebble predictor} considers a mono-disperse dust distribution at every radius, with solid, compact grains with size $a_{\rm{max}}(r,t)$, the size of which can change with time and radius and is set - after a brief early growth period - by either the fragmentation or radial drift limit. At $t=0$, the dust-to-gas ratio equals $Z_0 = \Sigma_\mathrm{d}/\Sigma_\mathrm{g} = 0.01$ everywhere and the dust distribution follows the self-similar gas density profile of Eq.~\ref{eqn:gas profile}. \texttt{pebble predictor} then tracks the Stokes number (and hence size) of particles in the disc as a function of time and radius through a radial grid of discretised bins. 
Fragmentation-limited particle growth is dictated by the dominant motion determining the relative velocities between particles (either turbulence- or drift-induced motion). In the case of turbulence-induced fragmentation, the maximum Stokes number that the particle can reach is given by:

\begin{equation}\label{eqn:turb frag grain size}
    \text{St}_{\text{frag,turb}} =f_{\text{f}}\frac{v_{f}^{2}}{3\alpha c_{s}^2}
\end{equation}
where $f_{\rm{f}}$ is a parameter set to 0.37, $v_{f}$ is the fragmentation velocity (which we set to 100 cms$^{-1}$ for silicate grains \citep{blum&wurm2008, birnstiel2009}, as in Table \ref{tab:model_parameters}), $\alpha$ is the turbulence parameter \citep{shakura&sunyaev1973}, and $c_{s}$ is the speed of sound in the gas \citep{drazkowska2021}. Similarly, the maximum Stokes number for radial-drift induced fragmentation is: 

\begin{equation}\label{eqn:drift frag grain size}
    \text{St}_{\text{frag,drift}}=f_{\text{f}}\frac{v_{f}}{2v_{\eta}}
\end{equation}
where $v_{\eta}$ is the maximum pebble drift speed, as determined by the gas, which is given as:

\begin{equation}
    v_{\eta} = \frac{1}{2 \rho_{\rm{gas}} \Omega_{K}} \cdot \frac{\partial P}{\partial r}
\end{equation}
for gas pressure $P$, midplane density $\rho_{\rm{gas}}$ and Keperlian orbital frequency $\Omega_{K}$.

When the particle growth is instead limited by the gas drag-induced inward radial drift of pebbles, the maximum Stokes number is given by:
\begin{equation}\label{eqn:drift grain size}
    \text{St}_{\text{drift}} = \frac{1}{f_{\text{d/g}} \eta} \cdot \frac{\Sigma_{\text{d}}}{\Sigma_{\text{g}}}
\end{equation}
for logarithmic gas pressure gradient $\eta$ and a factor $f_{\rm{d/g}}=30$ that dictates how many times faster particle growth is compared to drift \citep{okuzumi2012, drazkowska2021}. The radial drift velocity $v_r$ for any pebble is given as \citep{weidenschilling1977}:

\begin{equation}
    v_{r}=\frac{2 v_{\eta} \rm{St}}{1 + \rm{St}^{2}}
\end{equation}
which is limited by the fragmentation and drift barriers (equations \ref{eqn:turb frag grain size}, \ref{eqn:drift frag grain size}, and \ref{eqn:drift grain size}) by setting the maximum pebble size in these regimes. As the particles within the disc grow and begin to drift, mass within the disc begins to be redistributed in accordance with Eq.~\ref{eqn:pebble flux}.

The initial spatial particle size distribution is given via the equation for a particle's Stokes number in the Epstein drag regime. This is related related to its physical size $a$ through the equation:
\begin{equation}\label{eqn:epstein stokes}
    \text{St} = \frac{\pi}{2} \frac{a \rho_{s}}{\Sigma_{\rm{g}}}
\end{equation}
where density $\rho_{s}$ is the material density \citep[e.g.,][]{birnstiel2009}, which we set to $\rho=1.25 \mathrm{~g~cm^{-3}}$, following \citet{drazkowska2021}. For a typical disc, a millimetre-sized pebble will have $\mathrm{St} \sim 10^{-3}$ at 100 AU while a meter-sized one will be closer to $\mathrm{St} \sim 0.5$ (assuming $\Sigma_{\rm{g}}$ follows Eq.~\ref{eqn:gas profile}). 

Due to the aforementioned simplifying assumptions, \texttt{pebble predictor} has an extremely short runtime: a single run entailing the evolution of a dust disc takes around 30$\umu$s. This is the key advantage of using \texttt{pebble predictor} over a code that computes the full size distribution \citep[e.g. \texttt{DustPy,}][]{stammler&birnstiel2022};
the extremely short runtime means that \texttt{pebble predictor} lends itself very well to parameter-sampling methods such as Markov Chain Monte Carlo (MCMC) sampling.

\subsection{MCMC Priors and Likelihood}
\label{sec:MCMC_set_up}

To retrieve the initial disc parameters, such as the disc gas mass $M_{\rm{disc}}$ and characteristic radius $r_{c}$, from the inferred cumulative pebble flux, we used the MCMC sampler \texttt{emcee} \citep{foreman-mackey2013} in order to sample the simulation parameter space through a likelihood-based analysis. We sampled the parameter space using the logarithm of the disc gas mass, \logMdisc, and the logarithm of the characteristic radius, \logRcrit,  for a synthetic disc (section \ref{sec:test_case}) and HD 163296 (section \ref{sec:HD 163296}). We refer to the disc gas mass as the `disc mass' hereafter, unless otherwise specified. The priors we used were uniform for both parameters, representing a scenario with limited prior knowledge:

\begin{equation}
    \mathcal{P}_{\rm{prior}}=
    \begin{cases}
        1 & \text{for } -3 < \log(M/M_{\sun}) < -0.3 \\
          & \text{and } 1 < \log(r_{c}/\rm{AU})< 3\\
        0 & \text{otherwise}
    \end{cases}
\end{equation}

Our likelihood analysis used a simple $\chi^{2}$ fit to assess the fitting of the model: 

\begin{equation}\label{eqn:log_likelihood}
    \ln(\mathcal{L}) = -\frac{1}{2} \sum_{n=1}^{N}\left( \frac{M_{p,\text{sample}}-M_{p,\text{model}}}{\sigma_{M_{p}}} \right)^{2}
\end{equation}

\noindent where $M_{p,\text{model}}$ is the cumulative flux calculated from \texttt{pebble predictor} (using Eq.~\ref{eqn:cumulative flux}) given some sample parameters, $M_{p,\text{sample}}$ is the `target cumulative flux', and $\sigma_{M_{p}}$ the uncertainty. Only a single snowline was assessed for a cumulative pebble flux by \citet{zhang2020}, so we set $N=1$. 
We note that since the actual pebble flux through the CO snowline may be higher than reported by \citet{zhang2020}, our choice of likelihood function may not be appropriate. If the pebble flux is higher, the likelihood function would need to be wider or centered on a higher value. This would, in turn, increase the disc birth mass as a higher flux is required. We, however, choose this likelihood as it is an effective way to utilise the existing pebble flux estimate. The details of each \texttt{emcee} production run are given in section \ref{sec:test_case}.

\subsection{Test Case using a Synthetic Disc}
\label{sec:test_case}

\subsubsection{Set-up}\label{sec:synthetic_disc}

Before applying our framework to HD 163296, we perform a test to see to what extent the methodology outlined above can reproduce known disc parameters based on a cumulative pebble flux value. To this end, we constructed a synthetic disc with known parameters (listed in Table~\ref{tab:model_parameters}) and calculated the cumulative pebble flux through $r=35\mathrm{\ AU}$ at 5.3 Myr. The cumulative pebble flux at 5.3 Myr equalled $M_{p,\text{sample}} = 35.4~\mathrm{M}_{\earth}$, and an error of 10 $\rm{M}_{\earth}$ was assumed. 

\begin{table}
    \caption{Table containing the model parameters that were passed to \texttt{pebble predictor}. These were used to produce a cumulative pebble flux for \texttt{emcee} to match to for the synthetic disc. The method to determine $T_{0}$ for HD 163296 is outlined in section~\ref{sec:additional_set_up}.}
    \label{tab:model_parameters}
    \renewcommand{\arraystretch}{1.2}
    \begin{tabular}{ | l | ccccc |}
        \hline     
        Disc & $\alpha$ & $\gamma$ & $M_{\star}$ & $T_0$ & $v_f$ \\
         & & & ($M_{\sun}$) & (K) & (cms$^{-1}$) \\
        \hline
        Synthetic & $10^{-5}$ & 1 & 1.4 & 280 & 100 \\
        HD 163296 & $10^{-4}$ $^{(1)}$ & 0.8$^{(2,3)}$ & 2.0 $^{(3, 4, 5)}$ & 167.3 & $100{-}1000$ \\
        \hline
    \end{tabular}
    \textbf{References.} (1) \citet{powell2022}; (2) \citet{zhang2019}; (3) \citet{zhang2021}; (4) \citet{andrews2018}; (5) \citet{fairlamb2015}
\end{table}

\begin{table*}
    \caption{Table containing the reference values for \logMdisc, \logRcrit and cumulative pebble flux $M_{p}$ for the synthetic disc and HD 163296 model with different fragmentation velocities; each retrieved mass for HD 163296 is not impacted by the fragmentation velocity, as the CO snowline is in the drift-limited regime (see section \ref{sec:hd_vfrag}). We note that the reference values formed the basis for the initial MCMC guesses used during the `burn-in' phase.}
    \label{tab:results_table}
    \renewcommand{\arraystretch}{1.3}
    \begin{tabular}{ | l | c c c c c c |}
        \hline
         & & Reference values & &  Retrieved values & \\
        Disc & $v_{f}$ (cms$^{-1}$) & $\log(r_c/\rm{AU})$ & $\log(M_{\text{disc}}/M_{\earth})$ & $\log(r_c/\rm{AU})$ & $\log(M_{\text{disc}}/M_{\earth})$ \\
        \hline
        Synthetic & 100 & 2.2 & -1.8  & $1.98^{+0.67}_{-0.66}$ & $-1.66^{+0.40}_{-0.22}$\\
        HD 163296 & 100 & 2.22 $^{(1)}$ & -0.82 $^{(1,2)}$& $2.31^{+0.45}_{-0.46}$ & $-0.64^{+0.19}_{-0.24}$\\
            "     & 250 & " & " & $2.32^{+0.46}_{-0.47}$ & $-0.64^{0.19}_{-0.24}$\\
            "     & 500 & " & " & $2.32^{+0.45}_{-0.47}$ & $-0.64^{+0.19}_{-0.24}$\\
            "     & 1000 & " & " & $2.31^{+0.46}_{-0.46}$ & $-0.64^{0.19}_{-0.24}$\\
        \hline
    \end{tabular}
    \\
    \textbf{References.} (1) \citet{zhang2019}; (2) \citet{zhang2020}
\end{table*}

The only parameters that were sampled were \logMdisc and \logRcrit; all others remained fixed for all of the sample models. This is due to the fact that disc mass and radius are the most dominant parameters in determining the (cumulative) pebble flux (see Fig.~\ref{fig:PP_examples}).

Following guidance in \citet{foreman-mackey2013}, we used the chosen disc mass and characteristic radius to create a small, 2-dimensional `ball' of initial guesses for \texttt{emcee} to use; this `ball' was centred on the true disc's mass and radius values. 256 walkers were used in the MCMC sampling process and were allowed 128 `burn-in' steps - we emphasise that this burn-in phase results in a new distribution of walkers that is independent of the initial ball of initial guesses. This new distribution of walkers was then used to initialise another sampling process of 2500 steps in a production run, ensuring the chain was over 50 times the autocorrelation time. The resulting chain was trimmed at the start and then thinned by twice the autocorrelation time to reduce correlation between the Markov Chain steps and ensure that the walkers were properly sampling the posterior distribution. This final thinned chain provided a set of 4,000  guesses for \logMdisc and \logRcrit that could then be compared against the values used to produce the cumulative pebble flux, which are given in Tables~\ref{tab:model_parameters} and ~\ref{tab:results_table}.

\subsubsection{Results}
\label{sec:synthetic_disc_results}

\begin{figure}
	\includegraphics[width=\columnwidth]{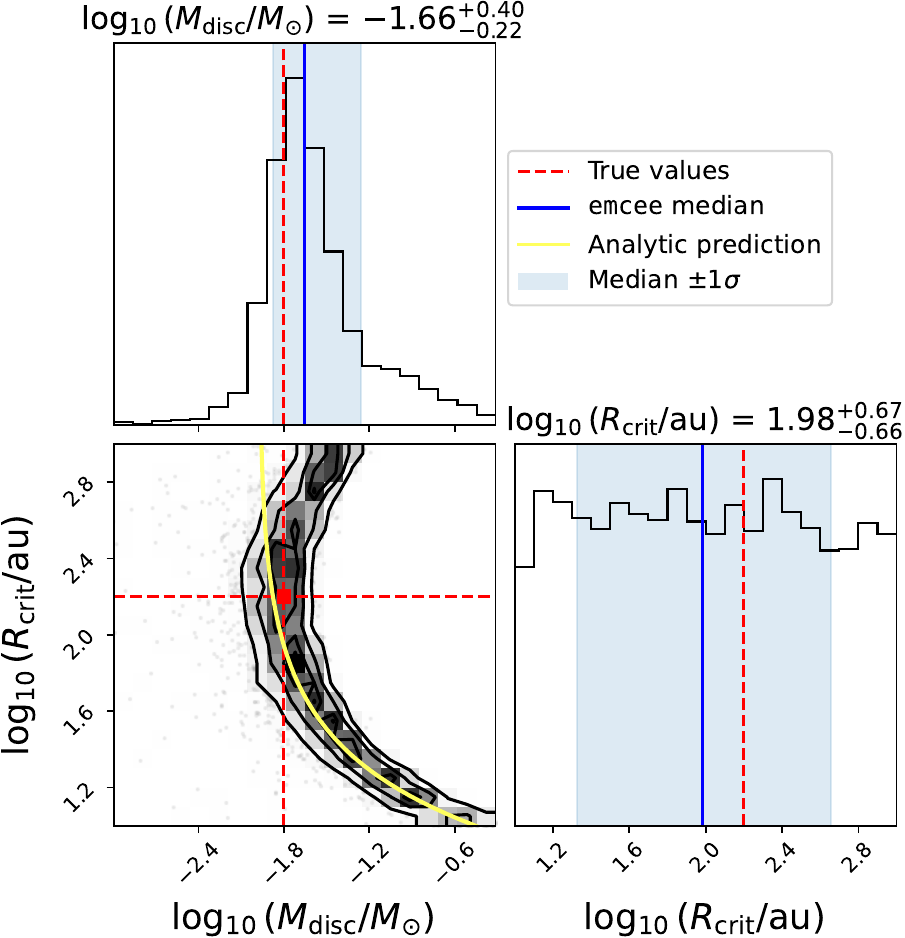}
    \caption{Corner plot of sample histograms for the synthetic disc of \logMdisc$=-1.8$ and \logRcrit$=2.2$. These `true' values are shown as the red dotted lines. The median parameter from \texttt{emcee}'s sample is shown as the solid blue line, with one standard deviation shown as the blue shaded area.}
    \label{fig:synthetic_disc_corner}
\end{figure}

The results of running emcee for the synthetic disc can be seen in the corner plot in Fig \ref{fig:synthetic_disc_corner}. The median values of the solutions accepted by the MCMC algorithm are indicated by the solid blue lines, with the median plus/minus one standard deviation on the results distribution shown with blue shading. The `true' values used to create the disc and target cumulative pebble flux are shown as the red dashed lines. The retrieved and true values are also reported in Table~\ref{tab:results_table}. \texttt{Emcee} finds values for both \logMdisc and \logRcrit that contain their reference values of -1.8 and 2.2 respectively. We note that the reported uncertainties are the 16th and 84th percentiles for the posterior distributions (lower and upper uncertainties respectively). The initial disc mass found by \texttt{emcee} has a relatively narrow standard deviation when compared to the characteristic radius; this is because the mass of the disc has a significantly larger impact on the cumulative pebble flux than the characteristic radius (see also Fig.~\ref{fig:PP_examples}). The masses reported in Table~\ref{tab:results_table} are independent of $v_{f}$ as the dust dynamics are drift-dominated outside the CO snowline, so the grain fragility is not a significant factor.

The posterior distribution for the disc radius is just the prior - that is, uniform - as the disc radius plays a much smaller role in determining the pebble flux in a disc compared to the birth mass. The disc radius therefore cannot be constrained with a single cumulative flux measurement.

The bottom-left panel in Fig.~\ref{fig:synthetic_disc_corner} highlights the degeneracy between the disc mass and characteristic radius. For example, solutions with very small discs (\logRcrit $\sim 1$) can be obtained but only for very high disc masses. This can be understood as follows: in the context of our disc model 
(that is, a static gas disc with pebbles that can only drift towards the central star), it is possible to find an analytic expression for the cumulative pebble flux through position $R$ in terms of the disc gas mass and characteristic radius in the limit where $t\rightarrow \infty$. Since all pebbles external to $R$ will have to pass through $R$ eventually (in the absence of planet formation or pressure traps), the cumulative flux after sufficiently long times will simply equal to the initial dust mass outside of $R$. 

For $\gamma=1$, the fraction of the disc's mass external to $R$ is given by (see Appendix~\ref{appendix:fractional mass derivation}):
\begin{equation}\label{eqn:fractional mass}
    f_{r\geq R} = \exp\left( -\frac{R}{r_{c}} \right)
\end{equation}

Using Eq.~\ref{eqn:fractional mass}, the mass of pebbles passing through $R$ is therefore:

\begin{equation}
    M_{p}(R, t\rightarrow \infty)=Z_0 M_{\text{disc}}\exp\left( -\frac{R}{r_{c}} \right)
\end{equation}

Rearranging for $M_{\text{disc}}$ yields an expression for the disc mass in terms of the characteristic radius:

\begin{equation}\label{eqn:banana equation}
    M_{\text{disc}} = \frac{1}{Z_0}M_{p}(R, t\rightarrow \infty) \exp\left( \frac{R}{r_{c}} \right).
\end{equation}

We then plot Eq.~\ref{eqn:banana equation} atop the 2D distribution of solutions in the bottom-left panel in Fig.~\ref{fig:synthetic_disc_corner} in yellow. This equation, and indeed \texttt{emcee}'s solutions, demonstrate that small discs (\logRcrit$\sim 1$) must be very massive to meet the required cumulative pebble flux.

Furthermore, that \texttt{emcee}'s accepted solutions follow this curve evidence the fact that it is finding physically plausible solutions. The spread around the curve is, of course, due to the fact that \texttt{emcee} is a parameter space sampler. At larger characteristic radii (\logRcrit$\gtrsim 2.5$), the solutions begin to diverge from Eq.~\ref{eqn:banana equation}. The reason for this is that very large discs evolve more slowly \citep{drazkowska2021}, and even after 5.3 Myr there is still a considerable dust reservoir outside of 35 AU. As a result, \texttt{emcee} tends to solutions with a higher dust mass then predicted based on Eq.~\ref{eqn:banana equation}, which assumes enough time has passed for $all$ pebbles from the outer disc to drift through $r=35~\mathrm{AU}$. 

In summary, trialling \texttt{emcee} on a synthetic disc of known parameters reveals interesting insights into existing degeneracies while highlighting its strength in retrieving the disc birth mass based on a single cumulative pebble flux measurement.

\section{HD 163296}
\label{sec:HD 163296}
\subsection{Set-Up}
\label{sec:additional_set_up}

We now turn our attention to HD 163296. As in Sect.~\ref{sec:synthetic_disc}, we focus on the disc gas mass and characteristic radius as the main free parameters. We set $\alpha=10^{-4}$ based on an (indirect) constraint based on vertical mixing timescales and sequestration of CO by \citet{Powell2019}, and $\gamma=0.8$ following \citet{zhang2019, zhang2020} (see Table~\ref{tab:model_parameters}). We set $v_f = 100\mathrm{~cms}^{-1}$ but investigate and discuss the impact of different (higher) values for the fragmentation velocity $v_f$ in Sect.~\ref{sec:hd_vfrag}. 

When fitting the cumulative pebble flux constraint, we decided to treat the estimated cumulative flux range as a data point with an associated Gaussian error 375±112.5 $M_{\earth}$ at a measurement time of 5~Myr based on the disc age \citep{montesinos2009, zhang2020}. In this way the corresponding range of 2$\sigma$ up or down corresponded to the minimum and maximum of the range reported in \citet{zhang2020} (that is, $150{-}600M_{\earth}$)

Since the cumulative flux estimate from \citet{zhang2020} is at the CO snowline (70 AU), we choose $T_0$ in Eq.~\ref{eqn:temperature profile} such that $T_{\rm{CO}}(70~\rm{AU})= 20~$K, corresponding to the sublimation temperature of CO \citep[e.g.,][]{zhang2015}. The resulting value for $T_0$ is shown in Table~\ref{tab:model_parameters}.

The MCMC sampling process was conducted identically to that for the synthetic disc described in Sect.~\ref{sec:synthetic_disc}. The final chain of results provided $\sim$9,700 solutions for $\log(M_\mathrm{disc}/M_{\odot})$ and \logRcrit.

\subsection{Results}
\label{sec:HD 163296_results}

\subsubsection{Disc Birth Gas Mass and Characteristic Radius}

\begin{figure}
	\includegraphics[width=\columnwidth]{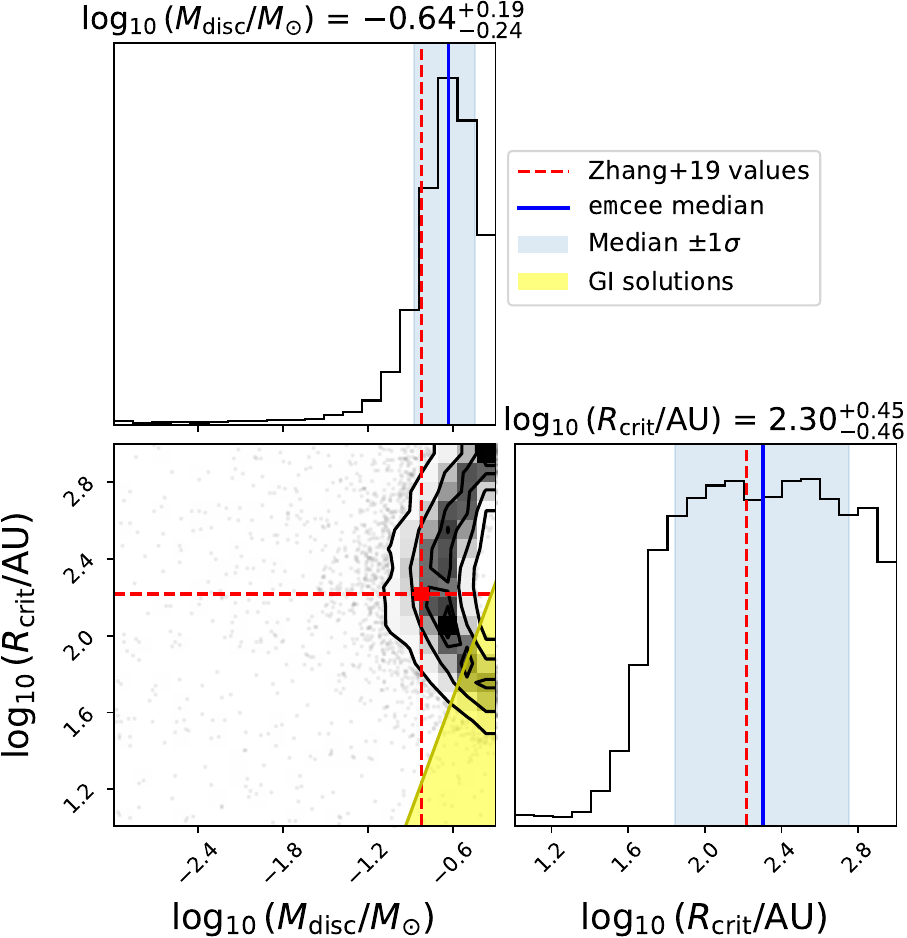}
    \caption{Corner plot of sample histograms for the model of HD 163296. Lines are the same as in Fig.~\ref{fig:synthetic_disc_corner}, but the `true' values instead correspond to observations from \citet{zhang2019} (red dotted lines). The yellow region contains discs that are gravitationally unstable.}
    \label{fig:HD 163296_corner}
\end{figure}

A target cumulative pebble flux of 375 ± 112.5 M$_{\earth}$ yielded the results that can be seen in Fig.~\ref{fig:HD 163296_corner}. Similarly to Fig.~\ref{fig:synthetic_disc_corner}, the median values from \texttt{emcee} are shown by solid blue lines, with one standard deviation around the median in the blue shading. The median disc birth mass is significantly higher than in the synthetic disc case. The characteristic radius is, again, degenerate, although smaller discs (\logRcrit$\lesssim 1.6$) are ruled out. This is because for \logRcrit$<1.85$, the amount of material available beyond 70 AU plummets rapidly, and it becomes very difficult to supply a sufficient quantity of pebbles through the CO snowline to match the target cumulative flux. This makes it very unlikely for discs with \logRcrit$\lesssim 1.6$ to be valid solutions. The characteristic radius posterior is therefore constrained only by the prior on the disc mass, by way of ruling out some disc sizes for certain disc masses, and is otherwise determined by the uniform prior. As such, our results for the characteristic radius are not constrained. Since there is a correlation between the mass of the disc and its radius (see Eq.~\ref{eqn:banana equation}), there is a small restriction on the mass of the discs. Whilst massive, compact discs remain possible, we calculated whether a disc was gravitationally unstable through the following equation \citep{kratter&lodato2016, stapper2024}:

\begin{equation}\label{eqn:grav_stability}
    \frac{M_{\rm{disc}}}{M_{\star}} > 0.06 \left( \frac{f}{1} \right) \left( \frac{T}{10~\rm{K}} \right)^{1/2} \left( \frac{r_{\rm{out}}}{100~\rm{AU}} \right)^{1/2} \left( \frac{M_{\sun}}{M_{\star}} \right)^{1/2}
\end{equation}

\noindent where $r_{\rm{out}}$ is the `outer radius' of the disc containing 90\% of the disc luminosity, and $T$ is the disc midplane temperature at $r$. Following \citet{stapper2024}, we set the prefactor $f=1$, and we approximated $r_\mathrm{out}$ by calculating the radius within which 90\% of the disc gas mass was contained. Discs that obeyed the inequality in equation~\ref{eqn:grav_stability} fall into the yellow region seen the lower-left panel in Fig.~\ref{fig:HD 163296_corner}, corresponding to massive compact discs that may not have survived 5 Myr; these were excluded from the solution set to produce a `filtered' set of solutions. 

Finally, we show in Fig.~\ref{fig:HD 163296_corner} the disc birth mass and characteristic radius as used by \citet[][Table~4]{zhang2019} to reproduce HD 163296's SED and radially resolved ALMA observations of $^{13}$CO and C$^{18}$O. These values are shown by the red dashed lines, and can also be found in Table~\ref{tab:results_table}. The agreement between the median solutions and the values of \citet{zhang2019} is remarkable, considering the only constraints that we provided to \texttt{emcee} was a single measurement of the cumulative pebble flux at one location (70 AU) and one time (5 Myr), alongside fixing other parameters. This agreement demostrates the efficacy of using pebble drift models to constrain disc birth conditions; the disc's bulk mass is intimately tied to the drift of pebbles and distribution of volatiles through the distribution of mass. The uncertainty in the characteristic radius is considerable (almost a factor of 10), and we discuss how this may be reduced by including cumulative pebble flux measurements of multiple locations (i.e. different snowlines) in Sect.~\ref{sec:discussion}. This is expected, however, for the disc mass impacts the pebble flux more significantly than disc radius. The disc mass obtained by our model is somewhat higher than the value reported by \citet{zhang2019} (although consistent to within $1\sigma$), but this can be understood in pointing out that our mass and that from \citet{zhang2019} are unlikely to be the same quantity: the disc mass we fit is the one at birth at $t=0$ whilst literature values like the ones of \citet{zhang2019} are based on current observations of the system at an age of $t \approx 5~\mathrm{Myr}$. Because our model does not evolve the gas disc mass (see Sect.~\ref{sec:methods}), a more involved model which includes viscous evolution and mass accretion may reduce this discrepancy.

Since the characteristic radius of the disc at birth is not able to be constrained with a single cumulative flux measurement, we focus on the disc birth mass for the remainder of this work.

\subsubsection{Evolution of the Cumulative Pebble Flux}
\begin{figure}
	\includegraphics[width=\columnwidth]{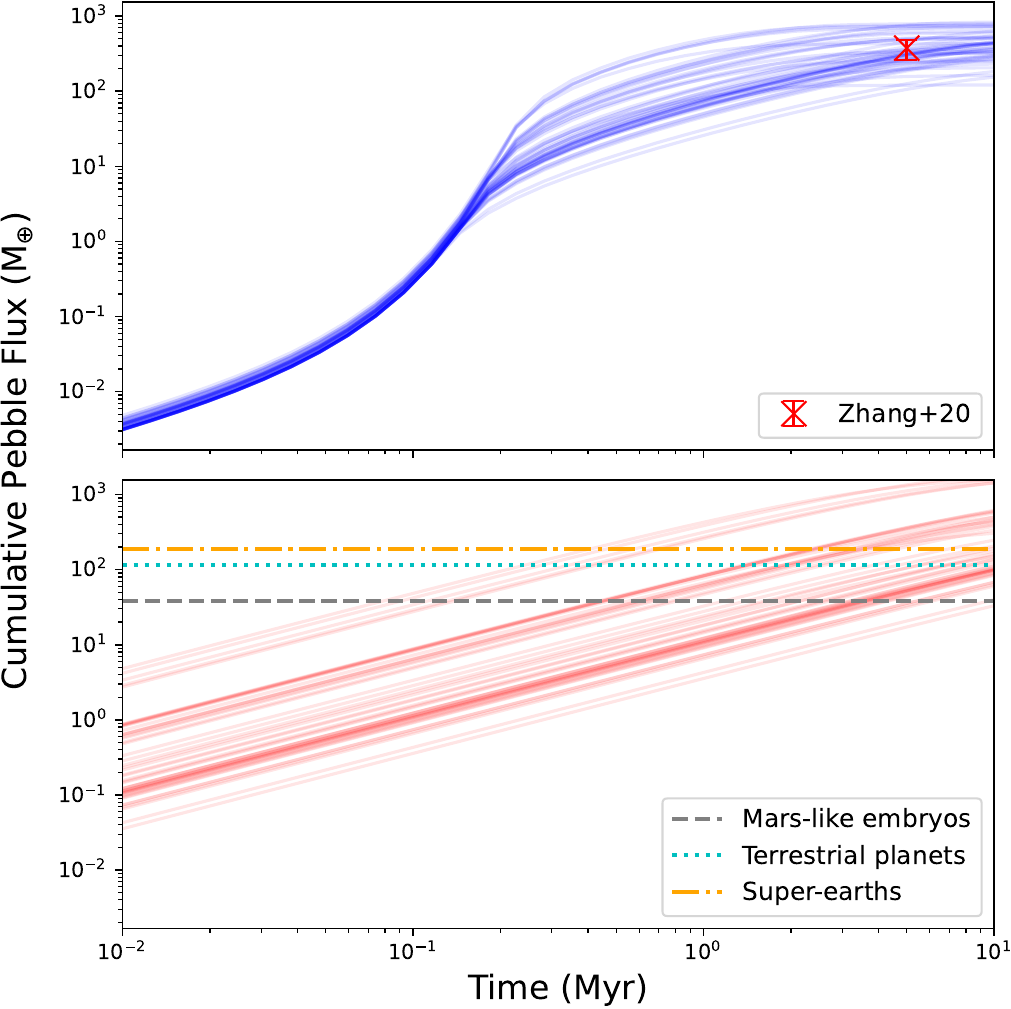}
    \caption{Cumulative flux at the CO snowline (70 AU, top) and water snowline (1.3 AU, bottom) as a function of time for 50 randomly selected \logMdisc and \logRcrit solutions from the posterior distribution. The target value derived from \citet{zhang2020} provided to emcee (375 ± 100 M$_{\earth}$) for sampling at the CO snowline is shown as the red cross. Cumulative fluxes from simulations by \citet{lambrechts2019} are shown for the water snow line, corresponding to a particular planetary architecture post disc-dispersal (grey, dashed for Mars-like embryos; cyan, dotted for terrestrial planets; and orange, dash-dotted for super-Earths).}
    \label{fig:snowlines_solutions}
\end{figure}

The top panel of Fig.~\ref{fig:snowlines_solutions} shows the cumulative flux at the CO snow line for fifty mass and radius solutions drawn from the posterior distribution of Fig.~\ref{fig:HD 163296_corner}. The target cumulative flux (375$\pm$112.5 $M_{\earth}$) is plotted on top for comparison. We can see that each of these lines follow a similar shape, with a rapid increase occurring around $0.2$~Myr before quickly following different paths. The most massive discs achieve a higher cumulative pebble flux, as demonstrated in Fig.~\ref{fig:PP_examples}. 

Since \texttt{pebble predictor} calculates the pebble flux at every location in the disc, we generated predictions for the total mass of pebbles crossing the water snowline (corresponding to 1.3 AU in our model, obtained from setting the temperature to the sublimation temperature of water, $T_{\mathrm{H_{2}O}}(r)=150~\mathrm{K}$, e.g. \citet{zhang2015}. We note that the water snowline in the disc around HD 163296 may in fact be further from the star \citep{notsu2019}. These predictions for the water snowline are shown in the bottom panel of Fig.~\ref{fig:snowlines_solutions}. Compared to the CO snowline solutions, the water snowline ones increase as a power law without plateauing - this is because not all of the dust exterior to the water snowline has passed through it, so the disc still continues to evolve. If the simulation were allowed to evolve for longer, then we would see these cumulative flux curves flatten out.

\begin{figure}
    \includegraphics[width=\columnwidth]{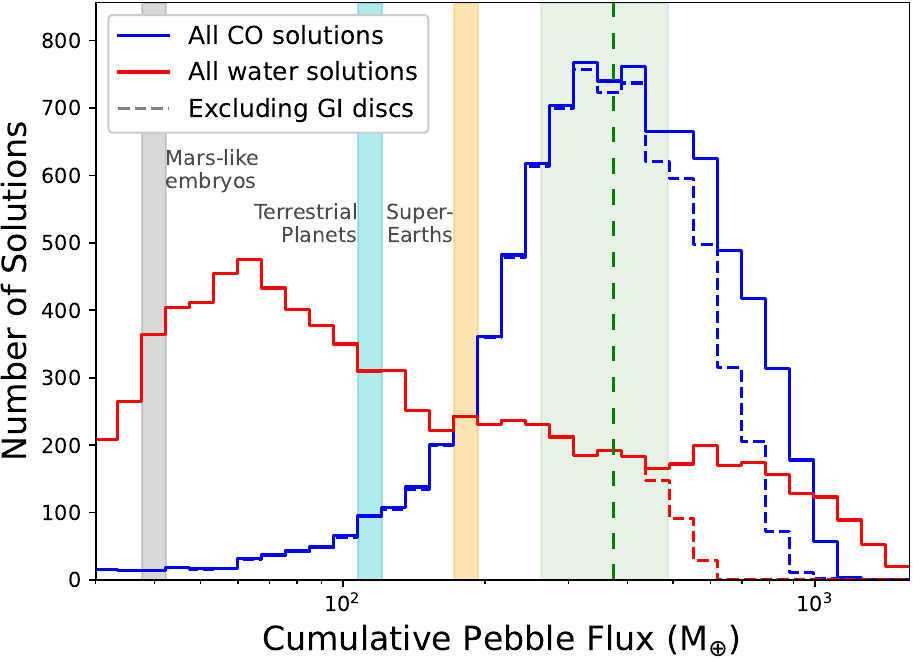}
    \caption{Histogram of the cumulative pebble flux through the CO (blue) and water (red) snowlines at 5 Myr, using all of the solutions from the posterior distribution (solid) and excluding gravitationally unstable discs (dashed). The target value derived from \citet{zhang2020} and provided to \texttt{emcee} (375 ± 112.5 M$_{\earth}$) is shown in green shading. As in Fig.~\ref{fig:snowlines_solutions}, cumulative fluxes from \citet{lambrechts2019} are shown, but as shaded bands to indicate which bin they correspond to.}
    \label{fig:snowlines_histogram}
\end{figure}

Figure ~\ref{fig:snowlines_histogram} shows a histogram of the cumulative pebble flux through the CO (blue) and water (red) snowlines at 5 Myr based on the full posterior distribution of solutions. The gravitationally unstable discs (see Eq.~\ref{eqn:grav_stability}) are filtered out in the dashed-line distributions, resulting in the higher cumulative pebble fluxes being filtered out. The distributions for CO peak around the target value: 40.4\% of the cumulative pebble fluxes as fall within the target range of $375\pm112.5$ M$_{\earth}$ (this is discussed further in section \ref{sec:discussion}). The water snowline distribution peaks at a lower cumulative pebble flux, primarily because 5 Myr is not long enough for all pebbles to have reached 1.3 AU.

Interestingly, the lower cumulative mass fluxes through the water snowline are, generally speaking, lower than those through the CO snowline. This is surprising, as more mass is contained exterior to the water snowline than the CO snowline. This, however, is less significant in our fiducial model with a fragmentation velocity of 100 cms$^{-1}$, since the dust is still growing and drifting at 5 Myr. If the fragmentation velocity is higher (e.g. 1000 cms$^{-1}$), then the water snowline \emph{does} have higher cumulative mass fluxes as the pebbles grow larger and drift more quickly. (see Appendix \ref{appendix:higher vfrag fluxes}). The impact of our choice of fragmentation velocity on the pebble flux in the disc can be seen in Appendix \ref{appendix: mulders flux plots}, where we see that the fragmentation velocity impacts the fragmentation-limited einner disc more than the drift-limited outer disc. This means that the cumulative mass flux through the CO snowline in the drift-limited region is insensitive to our choice of fragmentation velocity, whereas the water snowline in the fragmentation-limited region is more significantly impacted. We discuss the implications of our choice of fragmentation velocity on the dust mass evolution further in section \ref{sec:hd_vfrag}.

\subsubsection{Pebble Accretion and the Formation of Terrestrial Planets}
The water snowline predictions of Fig.~\ref{fig:snowlines_histogram} can be used to gain insights into plausible inner planetary system architectures by comparing them to the planet formation simulations of \citet{lambrechts2019}. These simulations use an exponentially decaying pebble flux reaching the inner disc until the disc is considered to be dispersed (at 10 Myr). Starting from a collection of planetary embryos, each simulation then followed their growth via pebble accretion, planet migration, as well as $N$-body interactions for 100 Myr after the protoplanetary disc dispersed. A clear connection was found between final planetary architectures and the cumulative pebbles flux to the inner disc during the first stages of the system's evolution \citep[see Fig.~7 of ][]{lambrechts2019}: cumulative fluxes of 38 M$_{\earth}$, 114 M$_{\earth}$ and $\geq$190 M$_{\earth}$ produced Mars-like embryos, terrestrial planets, and super-Earths respectively. We show these masses in the bottom panel of Fig.~\ref{fig:snowlines_solutions} and in Fig.~\ref{fig:snowlines_histogram} as shaded bands \footnote{We caution that the simulations by \citet{lambrechts2019} used a non-evolving Stokes number of $3\times 10^{-3}$ for all models. Since our models include varying Stokes numbers, we can expect different pebble accretion efficiencies of planetary embryos \citep{ormel&liu2018} and therefore possibly different system architectures to those predicted by \citet{lambrechts2019}.}. From Fig.~\ref{fig:snowlines_histogram}, we find that 12.8\% of the solutions predict planets smaller than Mars, 41\% predict ones between Mars-like embryos and Earth-like masses, 12.7\% predict between Earth-like and super-Earths, and 33.5\% predict super-Earth or larger planets. We note that the simulations of \citet{lambrechts2019} are for a sun-like star, whereas HD 163296 is not sun-like. The increased stellar mass would lead to a stronger gas pressure gradient, and therefore faster drift, perhaps delivering the required pebbles for planet formation sooner. We leave further study of the effect of stellar mass on planet formation outcomes to future work. It appears cumulative pebble fluxes between those producing Mars-like and terrestrial-mass planets are favoured in HD 163296, although Super-Earth-like systems cannot be ruled out.

\subsubsection{Dust Mass Evolution and Grain Fragility}\label{sec:hd_vfrag}

Our models so far have been based solely on fitting to the cumulative flux through the CO snowline and have assumed $v_f = 100~\mathrm{cm~s^{-1}}$. However, the composition of dust grains in HD 163296 is an open question \citep{guidi2022}, and so the exact value of the fragmentation velocity is also uncertain; this is because it depends on the composition and porosity of the grain material, which is currently not well constrained \citep[e.g.][]{blum&wurm2000, wada2013, weidling2012} which may vary with particle size and location. In this subsection we explore how estimates of the current dust mass can be used to rule out (much) higher values of the fragmentation velocity.

\begin{figure*}
	\includegraphics[width=0.8\paperwidth]{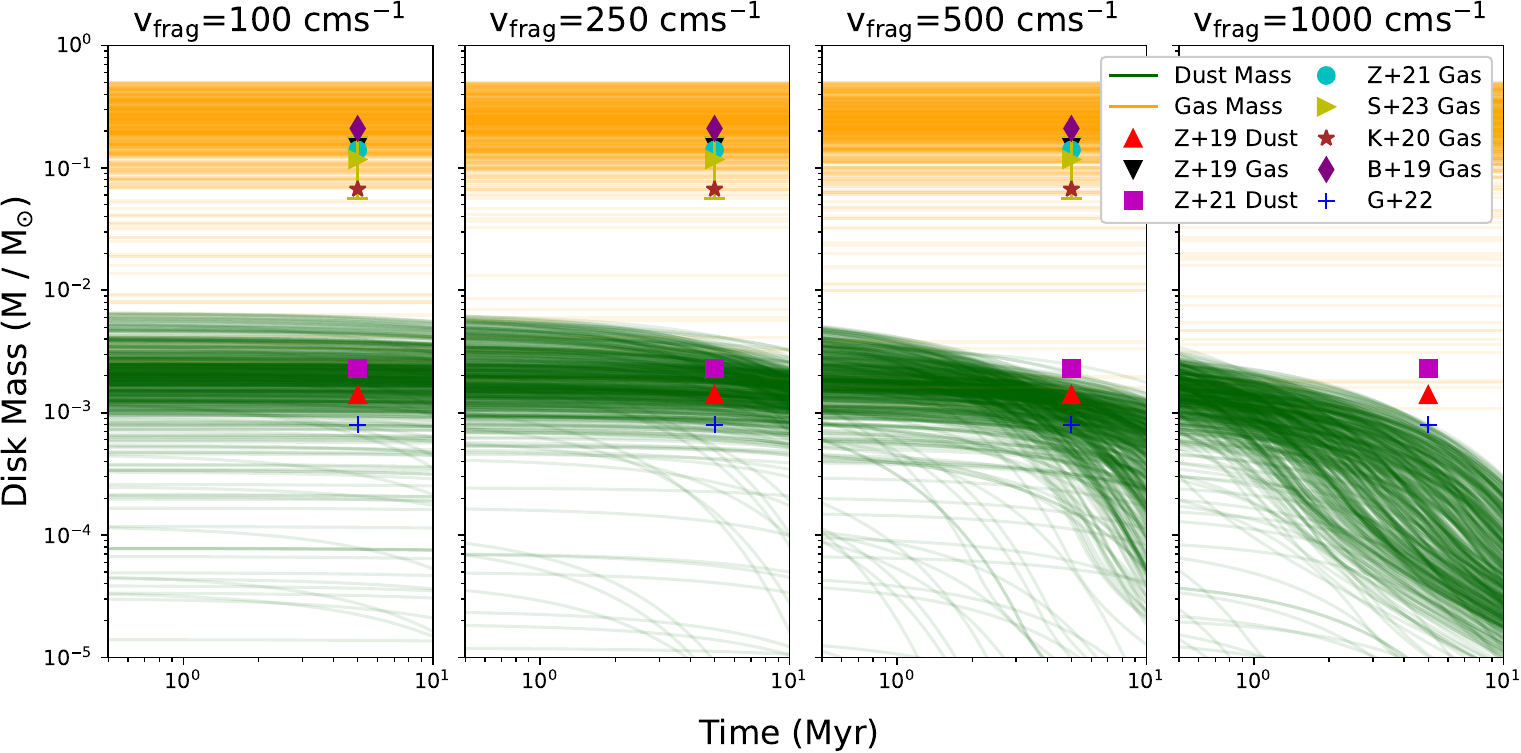}
    \caption{Dust and gas mass histories for 500 randomly selected values of \logMdisc and \logRcrit from four different \texttt{emcee} runs, each with different fragmentation velocities (A: 100 cms$^{-1}$, B: 250 cms$^{-1}$, C: 500 cms$^{-1}$, D: 1000 cms$^{-1}$). Each of the selected solutions were given to \texttt{pebble predictor} to calculate the dust and gas mass histories, although the gas disc is static in time. For comparison, the dust and gas masses from literature are shown (red and black triangles for \citet{zhang2019}; cyan square and magenta circle for \citet{zhang2021};  yellow triangle for \citet{stapper2022}; brown star for \citet{kama2020}; purple diamond for \citet{booth2019}; blue plus for \citet{guidi2022}. Lower fragmentation velocities inhibit pebble growth and cause particles to be less strongly influenced by the gas headwind, leading to significantly slower radial drift and dust reservoir depletion.}
    \label{fig:multiple_mass_histories}
\end{figure*}

To do this, we re-ran \texttt{emcee} for three additional fragmentation velocities (250 cms$^{-1}$, 500 cms$^{-1}$ and 1000 cms$^{-1}$), but otherwise the same parameters as described in Sect.~\ref{sec:HD 163296}. We then calculated the time evolution of the disc-integrated dust mass for 500 discs from the resultant mass and radius posterior distributions; these can be seen in Fig.~\ref{fig:multiple_mass_histories}, where each panel corresponds different fragmentation velocity. We compare these values against masses reported by \citet{zhang2019, zhang2021}, \citet{stapper2024}, \citet{kama2020}, \citet{booth2019}, and \citet{guidi2022}\footnote{Dust masses reported by \citet{zhang2021} and \citet{stapper2022} are based on ALMA (sub)millimeter continuum; \citet{zhang2021} use Band 3 and 6, and \citet{stapper2022} use Band 6 (1.29mm, see their Appendix C). \citet{kama2020} used the hydrogen deuteride rotational line emission, whereas \citet{booth2019} fitted models to azimuthally averaged profiles of optically thin $^{13}$C$^{17}$O isotopologue emission. \citet{guidi2022} integrated the dust surface density profiles returned from Monte Carlo fits of the intensity profile against data; this surface density was integrated between 8 and 120 AU \citep{guidi2022}. The authors additionally note that the calculated dust mass also depends strongly on the grains' compositions and their porosity, but both our model and the dust masses from \citet{guidi2022} shown in Fig~\ref{fig:multiple_mass_histories} assume that the grains are compact.}. We reiterate that our model does not evolve the gas mass nor surface density.

As shown in Table~\ref{tab:results_table}, the retrieved values of \logRcrit and \logMdisc are essentially independent of the choice of $v_f$; this is because the CO snowline sits in the drift-limited region of the disc (Eq.~\ref{eqn:drift grain size}), meaning the flux through the CO snowline is not impacted by the choice of fragmentation velocity. This leads to the same estimation for disc mass for different $v_f$. However, the \emph{timing} of when pebble drift occurs, and thus how much dust is left by 5 Myr, varies considerably as seen in Fig.~\ref{fig:multiple_mass_histories}.

\begin{figure}
	\includegraphics[width=\columnwidth]{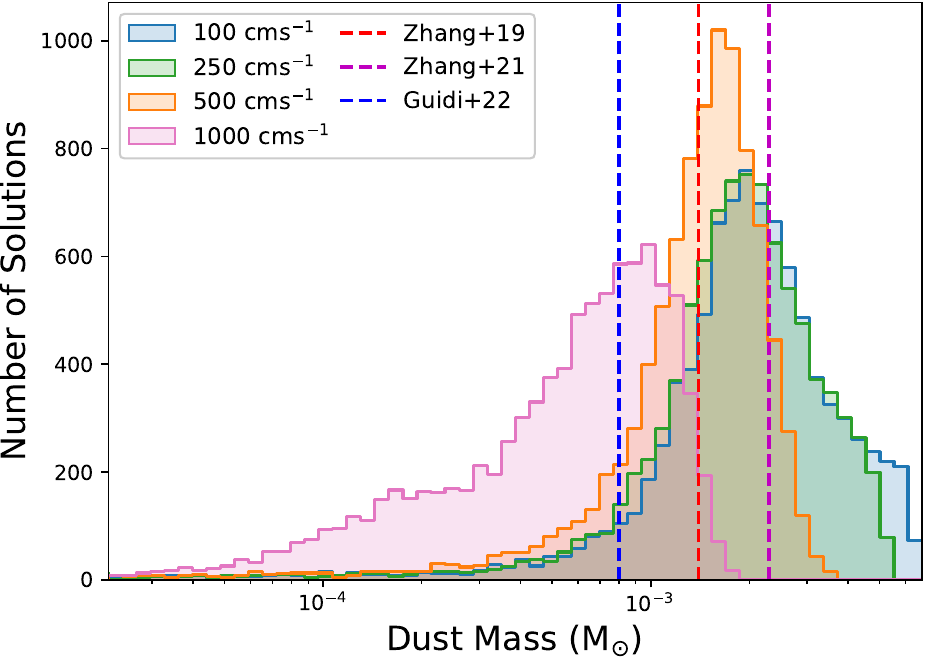}
    \caption{Histogram distributions of all disc dust masses at 5 Myr. Lower fragmentation velocities result in higher disc dust masses due to less pebble drift, causing the stronger peaks seen here. Based on visual inspection, the fragmentation velocity is very likely to be below 1000 cms$^{-1}$, suggesting the grains are fragile. However, the spread in dust masses between \citet{zhang2019, zhang2021} and \citet{guidi2022} makes it difficult to determine the best-fitting fragmentation velocity from this brief study.}
    \label{fig:mass_history_PDFs}
\end{figure}

It is evident that higher fragmentation velocities cause the disc dust reservoir to deplete more quickly. Since the maximum grain size is larger for higher fragmentation velocities in the fragmentation-limited regime (see equations \ref{eqn:turb frag grain size} and \ref{eqn:drift frag grain size}), pebbles drift faster, thereby shortening the drift timescale leading to the disc dust reservoir being depleted more quickly \citep{birnstiel2009, drazkowska2021}. \citet{stapper2024} report a dust-to-gas ratio of approximately 0.0014 for HD 163296, considerably lower than the canonical dust-to-gas ratio of 0.01, suggesting that the dust reservoir has depleted over time - this is consistent with our models of higher fragmentation velocities. A detailed view of the distribution of dust masses at 5 Myr for the full posterior distributions for each fragmentation velocity can be seen in Fig.~\ref{fig:mass_history_PDFs}. Again, increasing the fragmentation velocity causes a lower disc dust mass at 5 Myr, although increasing it significantly leads to the distribution poorly matching literature values on the current dust mass (as evidenced by the broad pink distribution for 1000 cms$^{-1}$).

We conclude then that while models with $v_{f} \geq 500 \ \rm{cms}^{-1}$ can accurately fit the cumulative mass flux through the CO snowline, pebble drift is so efficient in these simulations that these discs have very little dust left by 5 Myr, which is inconsistent with mm continuum observations \citep{zhang2019}. Conversely, models with relatively fragile dust grains ($v_{f} \leq 500 \ \rm{cms}^{-1}$) can match both the cumulative mass flux and remaining dust masses. Taking our fiducial model of $v_{f}=100$ cms$^{-1}$, the median dust mass was $\log(M_{\text{dust}}/M_{\sun})\ =-2.95^{+0.23}_{-0.24}$, which translates to $M_{\text{dust}}=0.0011^{+0.0008}_{-0.0005} M_{\sun}$ or $370^{+253}_{-156} M_{\earth}$, where the uncertainties are derived from the 16th and 84th percentiles of the dust mass distributions.

We note that our conclusion here is contingent upon a smooth disc model with no substructure, the presence of which would reduce the depletion rate of the dust reservoir (see section \ref{sec:substructure}).

\subsubsection{Radial Grain Size Distribution}

\begin{figure}
	\includegraphics[width=\columnwidth]{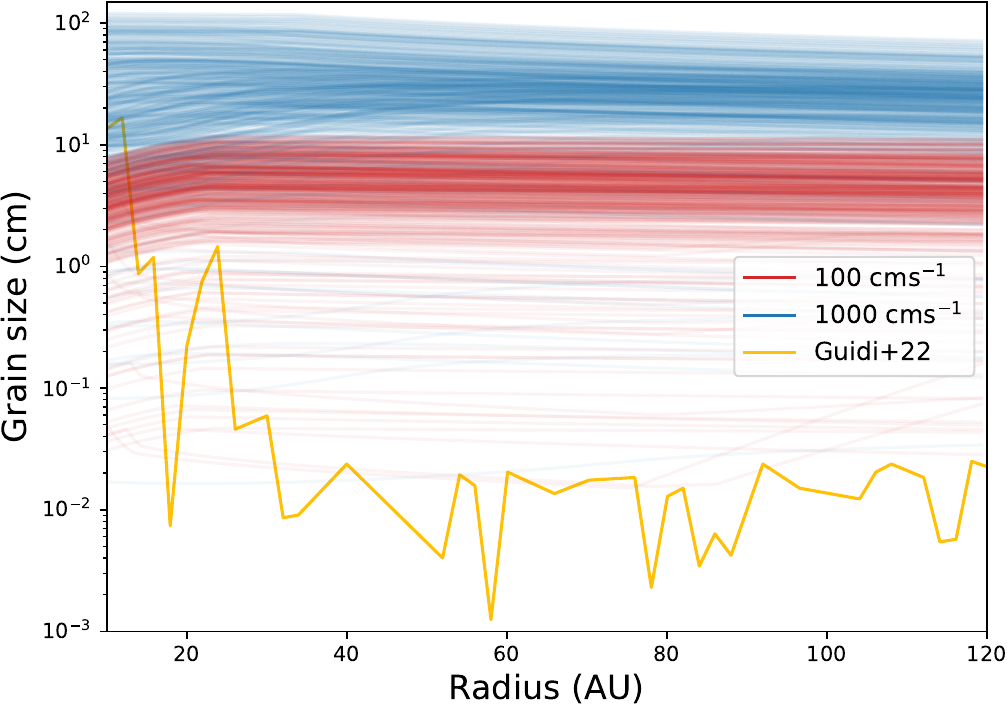}
    \caption{Grain sizes as a function of radius for 500 randomly selected solutions at 5 Myr for fragmentation velocities of 100 cms$^{-1}$ (red) and 1000 cms$^{-1}$ (blue). The maximum grain sizes of an example scattering model from \citet[][Fig.~9]{guidi2022} is shown in the single yellow line; the same radius range is shown for consistency.}
    \label{fig:grain_sizes}
\end{figure}

Finally, we conducted a brief comparison of the size of the solid, compact grains within our simulations against the distribution found by \citet{guidi2022}. We find that the distribution of grain sizes is approximately constant outside 30 AU, which is a consistent trend with results from models containing scattering by \citet{guidi2022}, although we find that our grains grown to larger sizes where the grain size is approximately constant (e.g. $\sim 5$ cm for $\rm{v_{f}}=100$ cms$^{-1}$ and $\sim$ 20 cm for $\rm{v_{f}}=1000$ cms$^{-1}$, compared  to 0.02 cm of \citet{guidi2022} at 70 AU). A plot showing the grain sizes for 500 random solutions at 5 Myr can be seen in Fig.~\ref{fig:grain_sizes}, with grain sizes from an example scattering model from \citet[][Fig.~9]{guidi2022} shown as a single yellow line. We note that the exact grain sizes depend on multiple factors, such as grain composition, porosity, and choice of scattering model. Nonetheless, the conclusion of approximately constant grain size outside $\sim$ 35 AU is independent of these factors, and sizes $\gtrsim 1~\rm{cm}$ are generally disfavoured in this region (see Appendix D of \citet{guidi2022}. We also emphasise that here we seek to compare the general trend of grain sizes, rather than attempting to constrain them to observations. In combining our dust mass comparison and our finding that lower fragmentation velocities produce dust particles closer in size to \citet{guidi2022} than higher ones, we conclude that lower fragmentation velocities are more likely to reproduce observations.

\section{Discussion}
\label{sec:discussion}

\subsection{Model Limitations and Extensions}
\label{sec:assumptions_and_limitations}

Our work is underpinned by several simplifying assumptions, most notably within \texttt{pebble predictor}. Since the MCMC process calls upon the simulation thousands of times, efficiency is a key factor in deciding which numerical model to use, and this led us to choosing \texttt{pebble predictor}, despite its simplicity. As a result, we have not included more complex mechanisms during the pebble growth and drift processes, such as the bouncing barrier \citep{zsom2010, guttler2010, dominik&dullemond2024}, or growth via vapour diffusion across snowlines \citep{ros&johansen2013}. We additionally do not evolve the gas density profile, nor does \texttt{pebble predictor} allow for the formation of dust traps and substructure formation. Substructure, however, has been observed within HD 163296 \citep{isella2016, huang2018}, and even the presence of exoplanets has been suggested as the cause of multiple dust gaps \citep{liu2018, teague2018, teague2021, pinte2018, guidi2018, izquierdo2022}. More comprehensive code packages, such as \texttt{DustPy} \citep{stammler&birnstiel2022}, are better suited to modelling discs containing substructure, but can take comparatively longer to run than \texttt{pebble predictor}. 

\subsubsection{Disc Substructure}
\label{sec:substructure}

The origins of disc substructure and their connection to ongoing planet formation processes are the subject of ongoing debate \citep[e.g.][]{dipierro2015, dong2015, dong2017, bae2017, teague2018, zhang_substructures2018, bae2018, liu2018}. What is clear, though, is that the presence of substructure in the gas disc can create local pressure bumps present at the outer edges of gaps that can slow down or the inhibit drift of large pebbles \citep{pinilla2012, drazkowska2019, kalyaan2021, sturm2022}. Despite this, the exact timing of substructure formation and their influence on the history of pebble drift in the disc is difficult to ascertain. Some young class-I discs feature substructures \citep{segura-cox2020}, whereas others have been observed to lack any notable features \citep{ohashi2023}. Even then, the formation of planets 1 Myr into the disc lifetime has a limited effect on the pebble trapping potential of pressure bumps \citep{stammler2023}, and substructures are required to form very early ($t<0.4$ Myr) to reproduce observed distributions of spectral index, disc size and luminosity \citep{delussu2024}, and even earlier ($t<0.1$ Myr) to significantly impact the inner disc water content \citep{mah2024}. If HD 163296 developed its current substructure features late into its evolution, then pebble drift could have occurred for a longer period, leading to a higher cumulative pebble flux, similar to what is believed to be the case for DL Tau \citep{sturm2022}. \citet{guidi2022} find that the size of dust grains does not change between the rings and gaps at 66 and 100 AU, indicating that the substructures could have formed relatively recently, or radial drift timescales are shorter than the ones associated with forming the substructures. On the other hand, the disc may have developed rings and gaps early on in its evolution, thereby significantly inhibiting the pebble drift and stalling out the growth of the cumulative flux early on \citep{mah2024}. \citet{drazkowska2016} and \citet{lambrechts&johansen2014} both note that since growth in the outer disc at early evolutionary stages is slower than the inner disc, there can be a delay in the delivery of material to the inner disc. If substructures form before the `pebble formation front' can reach the outer disc, then the outer disc could be cut off without having delivered any material. To then achieve an accurate cumulative pebble flux would require a significantly more massive disc.

In contrast, inhibiting pebble drift may be beneficial to the estimate of the disc dust mass. Scenarios with shorter pebble drift timescales (caused by larger $v_{f}$, see Fig~\ref{fig:multiple_mass_histories}) tend to undershoot the disc dust mass from literature at 5 Myr, so the inclusion of substructure may significantly lengthen this timescale and cause the dust mass history estimates to be larger at later times, despite larger fragmentation velocities tending to deplete discs more than is observed \citep{birnstiel2009}. This could cause the dust mass estimates to match more closely to observations, making higher fragmentation velocities more viable. In short, however, it is unknown how the cumulative flux could have evolved with time due to the impact of substructures.

Our constraint on the fragmentation velocity is additionally affected by substructures. The presence of a pebble trap would lead to slower dust reservoir depletion, which increases the remaining dust mass at later times. This would `lift up' the right-hand-side tails of the green dust mass curves in Fig.~\ref{fig:multiple_mass_histories}, and potentially bring larger fragmentation velocity models in line with current dust mass observations. This further highlights the importance of including substructure in tightly constraining dust properties such as grain fragility.

\begin{figure}
	\includegraphics[width=\columnwidth]{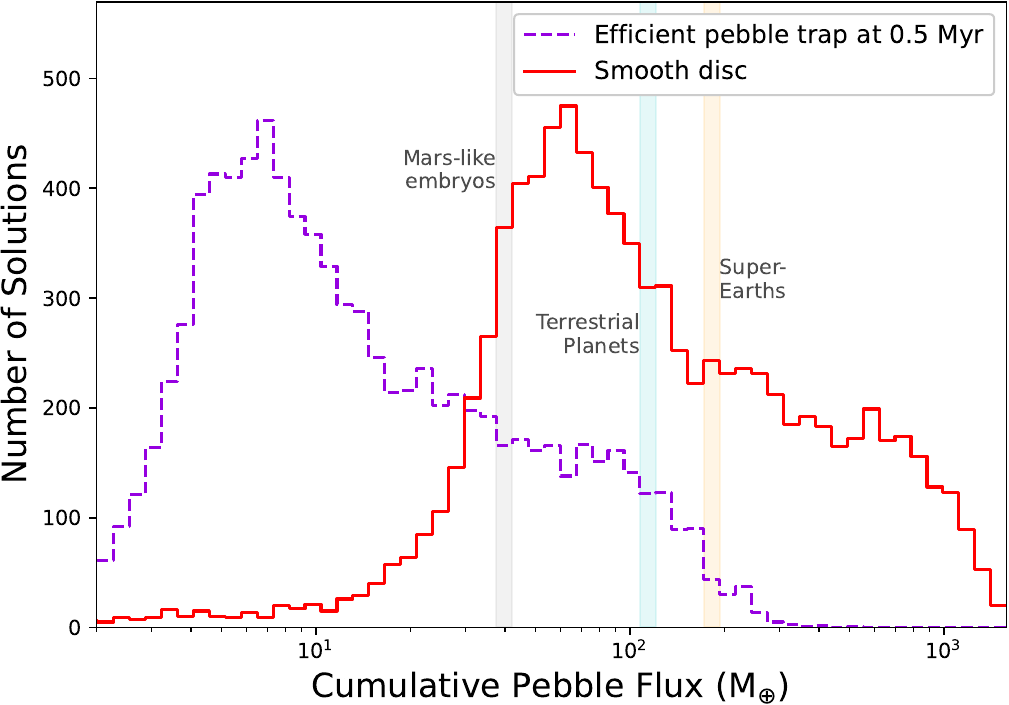}
    \caption{Prediction for the cumulative pebble flux through the water snowline by 0.5 Myr (purple dashed) and 5 Myr (red). The purple, dashed distribution represents a theoretically perfect pebble trap outside the water snowline and the red a smooth disc. Cumulative fluxes from \citet{lambrechts2019} are shown as bands, as in Fig.~\ref{fig:snowlines_histogram}, to demonstrate how the system's planetary architecture may change depending on when substructure-induced pebble traps form.}
    \label{fig:substructure_scenario}
\end{figure}

We briefly investigated the impact of a very efficient dust trap in HD 163296 by comparing the distribution of the cumulative pebble flux at the water snowline at 5 Myr (see Fig~\ref{fig:snowlines_solutions}) to the distribution at 0.5 Myr. The latter distribution represents a scenario where a gap and pressure bump forms (just) beyond the water snowline by 0.5 Myr and from then on completely blocks all further pebble drift into the region - a perfect pebble trap. This impact of this pebble trap can be seen in Fig~\ref{fig:substructure_scenario}. Since the cumulative pebble flux grows approximately as a power law in the bottom panel of Fig.~\ref{fig:snowlines_histogram}, it is unsurprising to see that cutting off the pebble drift earlier leads to a distribution of cumulative fluxes similar to the one for 5 Myr. In this pseudo-substructure scenario, ~78\% of the solutions lay in the Mars-like embryos region, ~17\% in `terrestrial planets', and ~5\% in super-earths. 

We note however that simulations of substructures demonstrate that they are not perfect dust traps, with some material passing through them due to fragmentation of larger pebbles. These `leaky dust traps' can lead to continued drift of smaller, fragmented pebbles through pressure bumps \citep[e.g.][]{rice2006, zhu2012, drazkowska2019, stammler2023}; the amount of material that would be allowed to pass through the substructures in HD 163296 is unknown, although other discs have been predicted to have dust traps with `trapping efficiencies' (i.e. how much material is inhibited from drifting) less than 100\% \citep{sturm2022}, meaning the results we provide here may yet remain accurate. A more detailed investigation of the impact of substructures will be the subject of future work.

\subsubsection{Further Assumptions}\label{sec:further_assumptions}

In this work, we assumed a disc age of 5 Myr based on \citet{zhang2020}, but the true age of the disc around HD 163296 is uncertain \citep{montesinos2009, gaia2018, zhang2020}, and the `age' of class-II discs can be difficult to define. If the stellar age is 5 Myr, then we might expect the class-II age to be lower. If the disc is indeed younger than 5 Myr, then the pebble flux through the CO snowline must be higher than predicted here in order to achieve the same cumulative flux; this would lead to a higher disc birth mass estimate based on the method we present here. The opposite is true if the disc is older than we assume. Substructure, however, complicates this picture - if the material is delivered entirely before the substructures appear ($t\lesssim 0.1$ Myr), then the disc age is less significant in our disc mass estimate.

Similarly, we assumed a cumulative flux value of 325$~M_{\earth}$ based on the work of \citet{zhang2020}. If the true cumulative flux value is at the higher end (600$~M_{\earth}$), then the disc estimate mass would be significantly higher; the opposite is true if the true cumulative flux is lower (150$~M_{\earth}$). Both our disc age and cumulative flux assumptions underpin our disc mass estimates, with a degeneracy between the two: older discs warrant lower masses, but larger cumulative fluxes require higher masses.

Additionally, we assumed the turbulence parameter $\alpha$ was constant throughout the disc, although \citet{liu2018} find that $\alpha$ is likely to vary throughout the disc (where it is lower in the inner disc compared to the outer disc). This was found in fitting the dust surface density profiles to observations via hydrodynamical simulations, and would consequently impact the dust mass of the disc. Our assumption of constant $\alpha$ would potentially impact the results we present here, although we assume a static self-similar surface density profile independent of $\alpha$, so our mass measurement is unlikely to change significantly.

Our $\chi^{2}$ likelihood function (Eq.~\ref{eqn:log_likelihood}) was symmetric around the sample cumulative flux $\rm{M_{p,sample}}$. Since the cumulative pebble flux is directly proportional to the disc mass, the distribution of cumulative pebble fluxes ought to be symmetric around the target value. \texttt{emcee}, however, was sampling parameters in logarithmic space as opposed to linear space, meaning that a small change $\Delta$\logMdisc$~$ at higher values of \logMdisc$~$ resulted in a significantly larger change in the cumulative flux than at lower values of \logMdisc. The consequence of this is that the resulting cumulative flux distribution appears to be positively skewed. This is simply a by-product of the sampling procedure and does not significantly impact the results.

\subsection{A Cumulative Pebble Flux Estimate in IM Lup}
Of the four systems studied by \citet{zhang2019}, HD 163296 was the only one for which the enhancement of CO interior to the CO iceline could be resolved (see their Fig.~8); \citet{zhang2021} found a similar enhancement in MWC 480. It is worth pointing out, however, that a cumulative pebble flux constraint exists for the younger IM Lup system as well. Using otherwise static thermo-chemical modelling, \citet{bosman2023} find that (at least) $540~M_{\earth}$ of dust needs to be added to the inner 20 AU of the disc in order to reproduce the observed depression seen in the radial profile of C$^{18}$O from Law et al. (2021). This suggests a very high pebble flux given IM Lup's young age (<1 Myr). These findings, combined with existing constraints on maximum grain size profiles \citep{Jiang2024} and turbulence levels \citep{Powell2019}, make IM Lup a promising case for applying our modelling approach, and MWC 480 also since it contains an enhancement similar to HD 163296; this will be the subject of a future study.

\subsection{Icy Pebble Drift across the Water Iceline with JWST}

Systems with pebble flux measurements at significantly different radial positions (e.g., at multiple different icelines) would be ideal test cases for constraining dust and pebble evolution models further. For example, \citet{bosman2018} investigated whether gas-phase CO$_2$ can be used to probe the efficiency of radial transport in across the CO$_2$ iceline (typically at ${\sim}10$ AU), and found that rapid CO$_2$ destruction may remove signatures of pebble drift (see their Fig.~7).

Closer to the star, tentative signatures of icy pebble drift across the H$_2$O snowline on a population-level were reported by \citet{banzatti2020}, based on combining Spitzer line fluxes and dust disc sizes from ALMA. More recently, \citet{banzatti2023} used JWST MIRI observations of four discs (two compact and two extended ones) to show that the differences in the water spectra seen in extended vs. compact discs stem largely from the presence or abundance of a `cold' water component that can be linked to icy pebble drift. Expanding on this idea, Muñoz-romero et al. (submitted) constrain the mass of this cold water reservoir for 7 nearby discs by fitting radially varying column density profiles to JWST MIRI spectra. Assuming the cold water in the disc surface is chemically processed on a timescale of ${\sim}100\mathrm{~yr}$, and that icy pebble drift is the sole supplier of this water, the pebble mass flux crossing the snowline can be estimated, with obtained values ranging from several ${\lesssim}30~M_{\earth}~\mathrm{Myr}^{-1}$ (for CI Tau) to $370~M_{\earth}~\mathrm{Myr}^{-1}$ (for FZ Tau) (Table 6 in Muñoz-Romero et al., submitted). Intriguingly, the lowest pebbles fluxes correspond to the most extended dust discs in the sample ($R_\mathrm{dust}> 100~\mathrm{AU}$), while the ones obtained for the more compact discs ($R_\mathrm{dust}\sim 10-20~\mathrm{AU}$) are an order of magnitude higher. 

Our estimates of the cumulative icy pebble flux through HD 163296's water snowline range from around $30~M_{\earth}$ to almost $10^3~M_{\earth}$ over 5 Myr (see our Fig.~\ref{fig:snowlines_histogram}), with the majority of solutions falling between $40 - 200~M_{\earth}$, or about $10 - 40~M_{\earth}~\mathrm{Myr}^{-1}$ when averaging over 5 Myr. This puts our predictions right in the middle of the values Muñoz-Romero et al. (submitted) obtained for extended dust discs - which appears to be consistent given HD 163296's large pebble disc ($R_\mathrm{dust} \approx 90~\mathrm{AU}$, from \citealt{huang2018}). 

\subsection{Radially Resolved CO Enhancements}
Our pebble flux constraint was based solely on the amount of CO enhancement interior to 70 AU as reported by \citet{zhang2020}. If available, however, the radial profile of this enhancement could be leveraged to place further constraints on the time evolution of the pebble flux (i.e., a recent high burst vs. a longer-lived but lower flux) and the efficiency of radial transport of gas-phase molecules trough advection and turbulent mixing.

For a constant pebble flux, \citet{cuzzi&zahnle2004} estimate that the gas-phase enhancement interior to an iceline will transition from being localized (their Regime 1, Fig.~3) to filling the entire inner disc (their Regime 2) on a timescale corresponding to roughly $(40/\alpha)$ orbital periods. For our disc model at 70 AU, this timescale considerably longer than the age of the system (even for a higher $\alpha=10^{-3}$). Models that treat dust coagulation, radial  drift, and CO redistribution self-consistently confirm this picture and find that lower values of $\alpha$ (i.e., weaker diffusion) lead to higher but more localized enhancements, while higher turbulence values tend to smear out the enhancement radially (see for example  Fig.~7 in \citealt{stammler2017}, and Fig.~11 in \citealt{krijt2018}). Whilst \citet{zhang2020} could not discriminate between a constant and a localized enhancement based on NOEMA observations, deeper high-resolution ALMA observations of CO may be able to provide such a constraint in the future. This is currently a subject under study (Armitage et al. in prep.),

\section{Conclusions}
\label{sec:conclusions}

In this work, we present a novel method of constraining the initial gas mass of the protoplanetary disc HD 163296. We combine fast dust coagulation and pebble drift calculations from \texttt{pebble predictor} \citep{drazkowska2021} with the MCMC sampler \texttt{emcee} \citep{foreman-mackey2013} to obtain cumulative pebble fluxes that passes through the CO snowline which we compare to values reported by \citet{zhang2020} based on an inferred gas-phase CO enhancement. Our main conclusions are as follows:

\begin{itemize}
    \item Through MCMC parameter space sampling, we determined the disc birth mass mass of HD 163296 to be \logMdisc $= -0.63^{+0.19}_{-0.24}$ ( $= 0.23^{+0.13}_{-0.10}~M_{\odot}$) and the characteristic radius of the disc to be \logRcrit $= 2.31^{+0.45}_{-0.45}$ ($=201^{+366}_{-131}$ AU; see Fig.~\ref{fig:HD 163296_corner}).
    \item Our models favour cumulative mass fluxes that suggest the disc around HD 163296 will produce planets that are between the size of Mars-like embryos and terrestrial planets, although we cannot rule out Super-Earth-like systems.
    \item We determine that dust grains are very likely to be fragile (fragmentation velocities of $v_f < 500 \mathrm{~cm~s^{-1}}$) as these reproduce currently observed dust masses better than more resilient grains (Fig.~\ref{fig:multiple_mass_histories}), based on our smooth disc model. 
    \item The radial distribution of the maximum grain size is approximately constant outside of 30 AU, which is a consistent trend with previous work. Our maximum grain size is, however, consistently larger than that found in literature.
    
\end{itemize}
Finally, we discuss the potential impact of disc substructure on our results and possible extensions of the approach presented here to other discs and/or other snowlines.

\section*{Acknowledgements}

The authors would like to thank the anonymous reviewer for their helpful comments and suggestions, which helped improve the manuscript. JW is funded by the UK Science and Technology Facilities Council (STFC), grant code ST/Y509383/1. The authors would like to thank Ke Zhang, Adrien Houge, Tilman Birnstiel, Sebastian Stammler, Joan Najita, Sebastian Marino, and Jenny Calahan for helpful suggestions and insightful comments, and Luke Deveson, whose MSc project helped inspire parts of this study. This work has made use of the Python packages \texttt{numpy} \citep{harris2020}, \texttt{matplotlib} \citep{hunter2007}, \texttt{scipy} \citep{virtanen2020}, and \texttt{emcee} \citep{foreman-mackey2013}.

\section*{Data Availability}

The data underlying this article will be shared on reasonable request to the corresponding author.

\bibliographystyle{mnras}
\bibliography{bibfile}

\appendix

\section{Fractional Mass Derivation}\label{appendix:fractional mass derivation}

The mass contained in a protoplanetary disc can be defined as:

\begin{equation}
    M_{\rm{disc}} = \int_{0}^{\infty} 2\pi r \Sigma_{g}(r) dr
\end{equation}

\noindent and hence the fraction of mass contained exterior to $R$ is:

\begin{equation}
    \begin{split}
        f_{r\geq R} = \frac{1}{M_{\rm{disc}}} &\int_{R}^{\infty}2\pi r\Sigma_{g}(r) dr
        \\
        = &\int_{R}^{\infty} r_{c}^{\gamma-2} r^{1-\gamma} \exp\left[ - \left( \frac{r}{r_{c}} \right)^{2-\gamma} \right] dr
    \end{split}
\end{equation}

\noindent Performing the substitution $x=(r/r_{c})^{2-\gamma}$ and recognising that $\lim_{r \rightarrow \infty} x = \infty$ for $\gamma<2$ yields:

\begin{equation}
    f_{r\geq R} = \int_{(R/r_{c})^{2-\gamma}}^{\infty} e^{-x} dx
\end{equation}

\noindent Which then, also undoing the substitution, evaluates to:

\begin{equation}
    f_{r\geq R} = \exp\left[ - \left( \frac{R}{r_{c}} \right)^{2-\gamma}\right]
\end{equation}

\noindent Which is the same as Eq.~\ref{eqn:fractional mass} when $\gamma=1$.

\section{Cumulative Mass Fluxes for Higher Fragmentation Velocity}\label{appendix:higher vfrag fluxes}

\begin{figure}
    \includegraphics[width=\columnwidth]{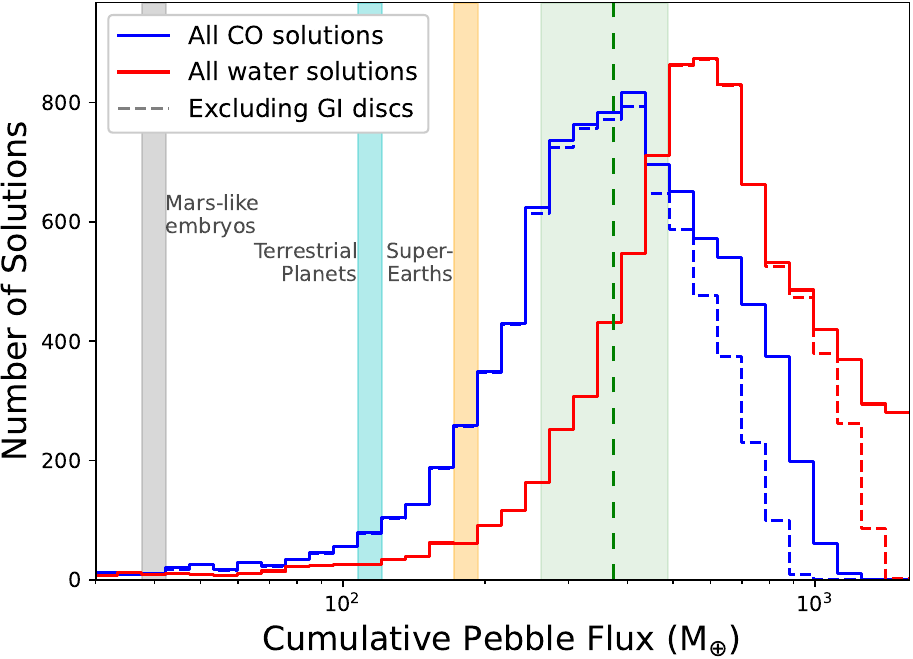}
    \caption{Same as Fig.~\ref{fig:snowlines_histogram}, but for a fragmentation velocity of 1000 cms$^{-1}$. The cumulative mass flux through the water snowline is larger than that through the CO snowline as the pebbles are able to grow to larger sizes and drift more quickly; this means that more material passes through the water snowline by 5 Myr compared to the fiducial $v_f=100~$cms$^{-1}$ case.}
    \label{fig:snowlines_histogram_vf_1000}
\end{figure}

\section{Impact of Fragmentation Velocity on Pebble Flux}\label{appendix: mulders flux plots}

\begin{figure}
    \includegraphics[width=\columnwidth]{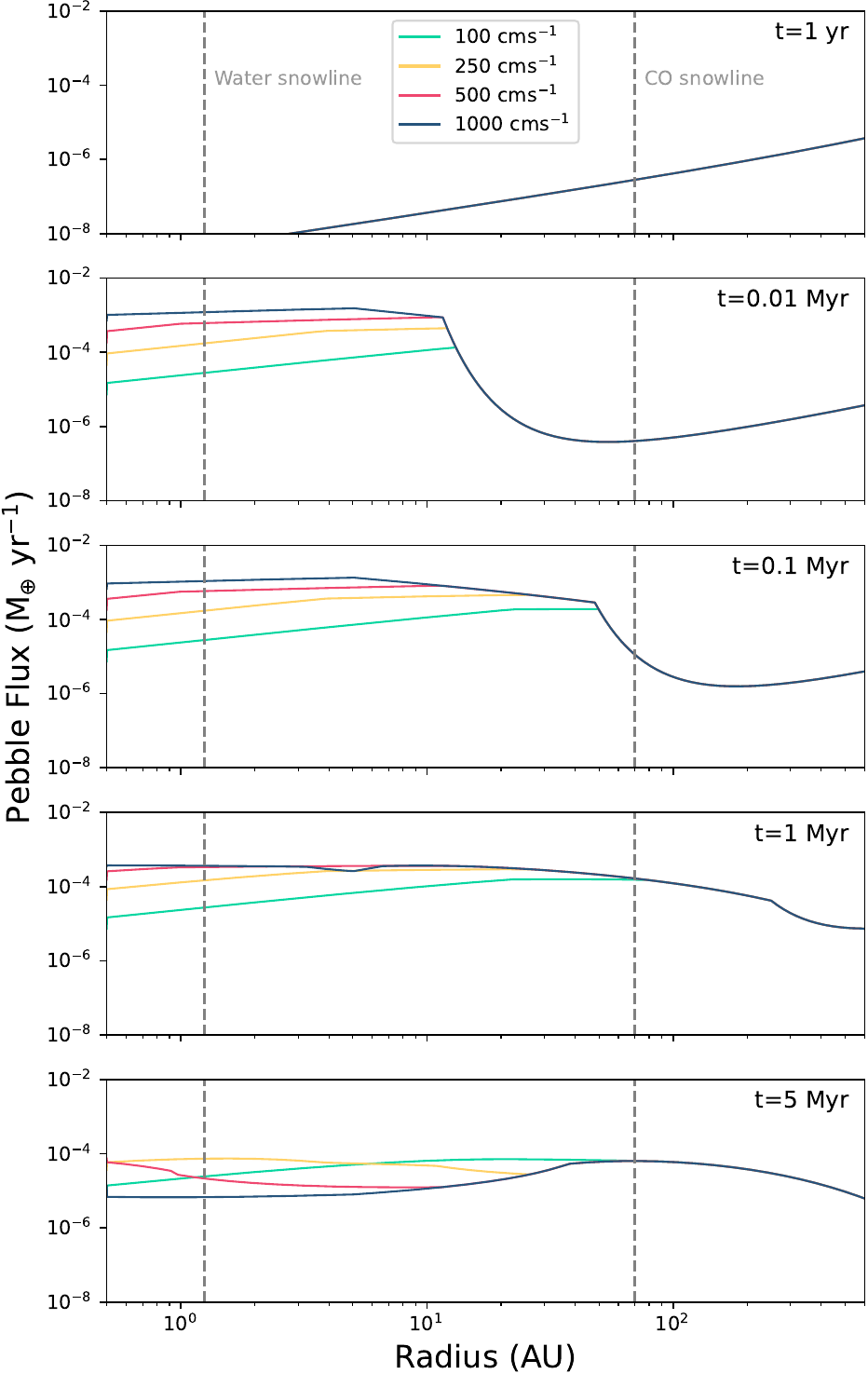}
    \caption{Pebble flux as a function of radius for a disc with the parameters of HD 163296 given in tables \ref{tab:model_parameters} and \ref{tab:results_table}, for different times and fragmentation velocities (denoted by different colours). Each disc achieves the same flux in the drift-limited region, where the CO snowline sits, but the fragmentation-limited region, where the water snowline is, shows different fluxes for different fragmentation velocities. This demonstrates that the mass flux through the CO snowline is insensitive to choice of fragmentation velocity.}
\end{figure}

\bsp
\label{lastpage}
\end{document}